\documentclass[%
 reprint,
 superscriptaddress,
 amsmath,amssymb,
 aps,
]{revtex4-2}

\usepackage{graphicx}
\usepackage{dcolumn}
\usepackage{bm}
\usepackage{algpseudocode}
\usepackage{ifthen}
\usepackage{pgfplots}
\usepackage[colorlinks=true,citecolor=blue,linkcolor=magenta]{hyperref}
\usepackage{ulem}
\usepackage{newfloat}
\usepackage{algcompatible}

\AtEndEnvironment{algorithm}{\par\vskip-5pt\noindent\hrulefill}

\DeclareFloatingEnvironment[
    fileext=loa,
    listname=List of Algorithms,
    name=ALGORITHM,
    placement=tbhp,
]{algorithm}

\newcommand{\Rulecomment}[1]{\hfill $\triangleright$ #1}

\algnewcommand{\IfThenElse}[3]{
  \State \algorithmicif\ {#1}\ \algorithmicthen\ {#2}\ \algorithmicelse\ {#3}}

\algnewcommand{\Repeat}[1]{\State \textbf{repeat} 
\ #1} \algnewcommand{\Until}[1]{\State \textbf{until} \ #1}

\begin{document}

\preprint{APS/123-QED}

\title{CrystalGRW: Generative Modeling of Crystal Structures with Targeted Properties via Geodesic Random Walks}

\author{Krit Tangsongcharoen}
\thanks{These authors contributed equally to this work.}
\affiliation{%
 Extreme Conditions Physics Research Laboratory and Center of Excellence in Physics of Energy Materials (CE:PEM), Department of Physics, Faculty of Science, Chulalongkorn University, Bangkok 10330, Thailand
}
\affiliation{%
Thailand Center of Excellence in Physics, Ministry of Higher Education, Science, Research and Innovation, 328 Si Ayutthaya Road, Bangkok 10400, Thailand
}
\author{Teerachote Pakornchote}
\thanks{These authors contributed equally to this work.}
\email[Correspondence to: ]{pakornct@mailbox.sc.edu}
\affiliation{%
 Department of Chemistry and Biochemistry, University of South Carolina, South Carolina 29208, USA
}
\author{Chayanon Atthapak}
\author{Natthaphon Choomphon-anomakhun}
\affiliation{%
 Extreme Conditions Physics Research Laboratory and Center of Excellence in Physics of Energy Materials (CE:PEM), Department of Physics, Faculty of Science, Chulalongkorn University, Bangkok 10330, Thailand
}
\author{Annop Ektarawong}
\affiliation{%
 Extreme Conditions Physics Research Laboratory and Center of Excellence in Physics of Energy Materials (CE:PEM), Department of Physics, Faculty of Science, Chulalongkorn University, Bangkok 10330, Thailand
}
\affiliation{%
Chula Intelligent and Complex Systems Lab, Department of Physics, Faculty of Science, Chulalongkorn University, Bangkok 10330, Thailand
}
\author{Björn Alling}
\affiliation{
Theoretical Physics Division, Department of Physics, Chemistry and Biology (IFM), Linköping University, Linköping, SE-581 83 Sweden
}
\author{Christopher Sutton}
\affiliation{%
 Department of Chemistry and Biochemistry, University of South Carolina, South Carolina 29208, USA
}
\author{Thiti Bovornratanaraks}
\email[Correspondence to: ]{thiti.b@chula.ac.th}
\affiliation{%
 Extreme Conditions Physics Research Laboratory and Center of Excellence in Physics of Energy Materials (CE:PEM), Department of Physics, Faculty of Science, Chulalongkorn University, Bangkok 10330, Thailand
}
\affiliation{%
Thailand Center of Excellence in Physics, Ministry of Higher Education, Science, Research and Innovation, 328 Si Ayutthaya Road, Bangkok 10400, Thailand
}
\author{Thiparat Chotibut}
\email[Email: ]{thiparatc@gmail.com}
\affiliation{%
Chula Intelligent and Complex Systems Lab, Department of Physics, Faculty of Science, Chulalongkorn University, Bangkok 10330, Thailand
}

\date{\today}

\begin{abstract}
Determining whether a candidate crystalline material is thermodynamically stable depends on identifying its true ground-state structure, a central challenge in computational materials science. We introduce CrystalGRW, a diffusion-based generative model on Riemannian manifolds that proposes candidate crystal configurations in stable phases, validated through density functional theory calculations. The crystal properties, such as fractional coordinates, atomic types, and lattice matrices, are represented on suitable Riemannian manifolds, ensuring that new predictions generated through the diffusion process preserve the periodicity of crystal structures. We also incorporate an equivariant graph neural network to account for rotational and translational symmetries within the model. CrystalGRW generates crystal structures that are stable and closely resemble their density‑functional‑theory ground states. The model also supports conditional control, such as enforcing a specified crystallographic point group, thereby accelerating materials discovery and inverse design by providing symmetry‑consistent, energetically stable candidate crystals for experimental validation.
\end{abstract}

\maketitle


\section{\label{sec:intro}Introduction}

Crystal structures, defined as the periodic arrangement of atoms in a lattice, directly influence material properties such as stability, band gap, and mechanical strength. 
Understanding and predicting the behavior of materials from the atomistic level relies on knowing accurate structures. The search for novel materials involves identifying crystal structures and their compositions that are thermodynamically metastable. 
However, this task is complicated by the vast number of degrees of freedom that must be explored to identify local minima on the free energy surface ~\cite{Oganov2019-dw, Needs2016}.

Many advanced algorithms for sampling configurations on the potential energy surface have been  developed for ground state structure prediction, these approaches typically combine (stochastic) samples technique such as evolutionary algorithms, \textit{ab-initio} random structure searching, simulated annealing, metadynamics, basin hopping, and molecular dynamics simulations \cite{Needs2016, Uspex2006, Calypso2010, Kirkpatrick1983simulatedannealing, Laio2002metadynamics, Wales1997basinhopping, CarParrinello1985aimd}, with quantum mechanics-based method, such as density functional theory (DFT) to  accurately predict the energy of the sampled configurations.  
However, the cubic scaling with respect to the number of basis functions in DFT leads to a substantial computational time for searching for new candidate materials. Machine-learned interatomic potentials (MLIPs) are an accurate and efficient alternative to DFT~\cite{Batatia2024mace, Deng2023chgnet, Chen2022megnet, Park2024sevennet, Liao2023equiformer, Liao2024equiformerv2, Bochkarev2024grace, Timmerman2024otfmlmd, Karimitari2024hoip}, and therefore can speed up the prediction process. Nonetheless, improving the structure generation task remains an important challenge for materials discovery.

Instead of the traditional approach of starting from an initial structure and performing sampling to find new (possibly lower energy) structures, generative models enable a direct and much more efficient way to produce novel molecular and crystal structures because they can sample new materials given knowledge from the learned data distribution without being constrained by energy barriers \cite{Xu2021confgf, Xu2022geodiff, Guan2023edm, Xu2023geoldm, Xie2022cdvae, Pakornchote2024dpcdvae, Jiao2023diffcsp, Jiao2024diffcsppp, Miller2024flowmm, Sriram2024flowllm, Zeni2025mattergen}. Diffusion-based generative models draw principles from non-equilibrium thermodynamics to transform random noise into structured data \cite{Sohl-Dickstein2015, Ho2020, Song2019, Song2021, Karras2022elucidating}. These models generate samples through stochastic differential equations (SDEs) that consist of data-dependent drift and diffusion terms \cite{Anderson1982, Follmer1985, Cattiaux2023, Girsanov1960, Karatzas1998, Arnaudon2014browniangeo}. The Brownian motion component introduces stochastic noise, enabling the  generation of new crystal structures.  
Recent works show that properly designed diffusion models can generate crystal structures with realistic characteristics, such as atoms do not overlap, lattice parameters are reasonable, and crystallographic symmetries are imposed \cite{Xie2022cdvae, Pakornchote2024dpcdvae}.

Moreover, materials discovery focuses not only on generating new structures but also on their properties. For subsequent property prediction, the candidate structures produced from generative approaches tend to closely resemble the DFT-optimized structures (similar to performing a structure relaxation with MLIPs), thereby reducing the time required for subsequent DFT relaxations. Additionally, generative models can be applied for inverse design, where the generated materials are tailored to possess specific, controlled properties \cite{Zeni2025mattergen, Takahara2024generativeinversedesigncrystal, Guo2024diffusionxrd, Riesel2024genpxrd, Choudhary2024atomgpt, Luo2025crystalflow}.

To specify crystal structures, three properties are essential: fractional coordinates, atomic types, and lattice matrices. The first two properties reside in distinct non-Euclidean spaces, while the third property lies in Euclidean space.
In particular, fractional coordinates lie on the 3D torus \(\mathbb{T}^3 = [0,1)^3\) (Appendix \ref{sec:torus}), and atomic‐type compositions lie on the \(d\)‐dimensional simplex \(\Delta^d\), with \(d \sim 100\) in many inorganic materials (Appendix \ref{sec:hypercube}). Introducing naive stochastic noise can corrupt these properties by moving them outside their natural domains. This issue can be circumvented by introducing noise distributions that align with their domains, such as wrapped normal distributions for fractional coordinates and categorical distributions for atomic types \cite{Jing2022torsional, Austin2021categorical}. Models like DiffCSP \cite{Jiao2023diffcsp} and MatterGen \cite{Zeni2025mattergen} employ these domain-aligned approaches to generate crystal properties, thereby improving crystal structure generation.

Instead of selecting specific noise distributions that align with the natural domains of the properties, Bortoli et al. \cite{Bortoli2022rsgm} introduced the Riemannian Score-Based Generative Model (RSGM), which extends diffusion models to operate on Riemannian manifolds. In RSGM, data undergo random walks along geodesic paths on manifolds that appropriately represent the data's natural domain, rather than diffusing strictly within Euclidean space. This framework enables the use of manifolds that accommodate non-Euclidean crystal properties, such as 3D torus for fractional coordinates and $d$-simplex for atomic types. Recent advances have further enhanced diffusion models on Riemannian manifolds. Riemannian flow matching \cite{Chen2024rfm} unifies diffusion and flow models within Riemannian manifolds, while FlowMM \cite{Miller2024flowmm} leverages this approach using optimal transport-based distributions for crystal structure generation, thereby improving generation performance and accelerating the sampling process.

In this work, we introduce CrystalGRW, a generative model designed for crystal structure generation with specifiable target properties. CrystalGRW combines the Riemannian score-based generative model (RSGM) with EquiformerV2 \cite{Liao2024equiformerv2}, an equivariant graph neural network with an attention mechanism. The model generates three key properties of crystals: atomic coordinates, atomic types, and lattice parameters. We demonstrate that CrystalGRW can produce realistic crystal structures that are close to their ground states with accuracy comparable to existing models. Moreover, our model incorporates conditional control to specify properties of the generated structures using classifier-free guidance \cite{Ho2021cfg}. To illustrate this conditional control capability, we train CrystalGRW to generate structures with specific point groups as dictated by the input conditions, highlighting its potential for targeted materials discovery.

\section{Results and Discussion}
\label{sec:results}
\subsection{CrystalGRW}
The crystal generative model aims to generate crystal properties $\mathbf{x} \in \left\{\mathbf{\tilde r}, \mathbf{\tilde a}, \mathbf{L}\right\}$ of $N$ atoms in a unit cell where $\mathbf{\tilde r} =  \left(\mathbf{r}^{(1)}, \dots, \mathbf{r}^{(N)}\right)$ are fractional coordinates of each atom, $\mathbf{\tilde a} = \left(\mathbf{a}^{(1)}, \dots, \mathbf{a}^{(N)}\right)$ are their atomic types, and $\mathbf{L}$ is a lattice matrix. Similar to diffusion probabilistic models, we generate new data by reversing a diffusion process, wherein trained neural networks map noise distributions to structured data distributions. In this work, new crystals emerge through geodesic random walks on manifolds that capture each property: a 3D torus for fractional coordinates \(\mathbf{r}^{(i)}\), a \(d\)-dimensional simplex \(\Delta^d\) for atomic‐type probabilities \(\mathbf{a}^{(i)}\), and Euclidean space for the lattice matrix \(\mathbf{L}\). See Section~\ref{sec:RSGM} and Appendix~\ref{appendix:manifolds} for details.

Although atomic‐type vectors naturally reside in \(\Delta^d\), directly performing random walks on the simplex can be cumbersome due to 
boundary conditions of its geometry. To circumvent this, we adopt a uniform‐spacing {\it bijection} \cite{Devroye2013-uniform-spacing} that maps \(\mathbf{a}^{(i)} \in \Delta^d\) to \(\mathbf{A}^{(i)} \in \mathbb{C}^d\), \(d\)-dimensional hypercube. We then carry out the geodesic random walk on \(\mathbf{A}^{(i)} \in \mathbb{C}^d\), where reflections at the boundary are straightforward (see Eq.~\eqref{eq:hyper_expm} and Appendix~\ref{sec:hypercube}), and finally map back to \(\mathbf{a}^{(i)}\in\Delta^d\). This approach preserves the simplicity of a hypercube‐based walk while respecting the simplex geometry of atomic types.

For denoising crystal properties in reverse diffusion, we employ EquiformerV2 \cite{Liao2024equiformerv2}, an {\it equivariant graph neural network} (GNN) that accounts for rotational and translational equivariance to effectively capture tensorial and vectorial quantities, e.g. atomic forces \cite{Cohen2018sphericalcnn, Tomas2018tfn, Liao2023equiformer, Passaro2023escn}. Specifically, EquiformerV2 uses \(l\)-degree irreducible representations (irreps) of the \(\mathrm{SO}(3)\) group \cite{Geiger2022e3nn}. 
Fractional coordinate denoiser is produced via an \(l=1\) output head that yields a 3D equivariant vector, while atomic type and lattice matrix denoisers come from \(l=0\) output heads providing 100- and 9-dimensional invariant features, respectively. Further details are given in Appendix~\ref{appendix:equiformerv2}.

Fig.~\ref{fig:model} illustrates training and sampling schemes of CrystalGRW. During training, crystal properties are corrupted using the geodesic random walk (GRW), as described in Algorithm~\ref{algo:forward_grw}. A distorted structure is then constructed from the corrupted properties and passed through the graph neural network to learn the score function. For crystal structure generation, the three crystal properties are initially sampled from a random distribution and gradually denoised through a \emph{reverse} GRW process, computed using the learned scores.
When using the condition-guided scheme, specific conditions are embedded alongside the node features during both the training and generation processes.

\begin{figure*}[t] \includegraphics[width=\linewidth]{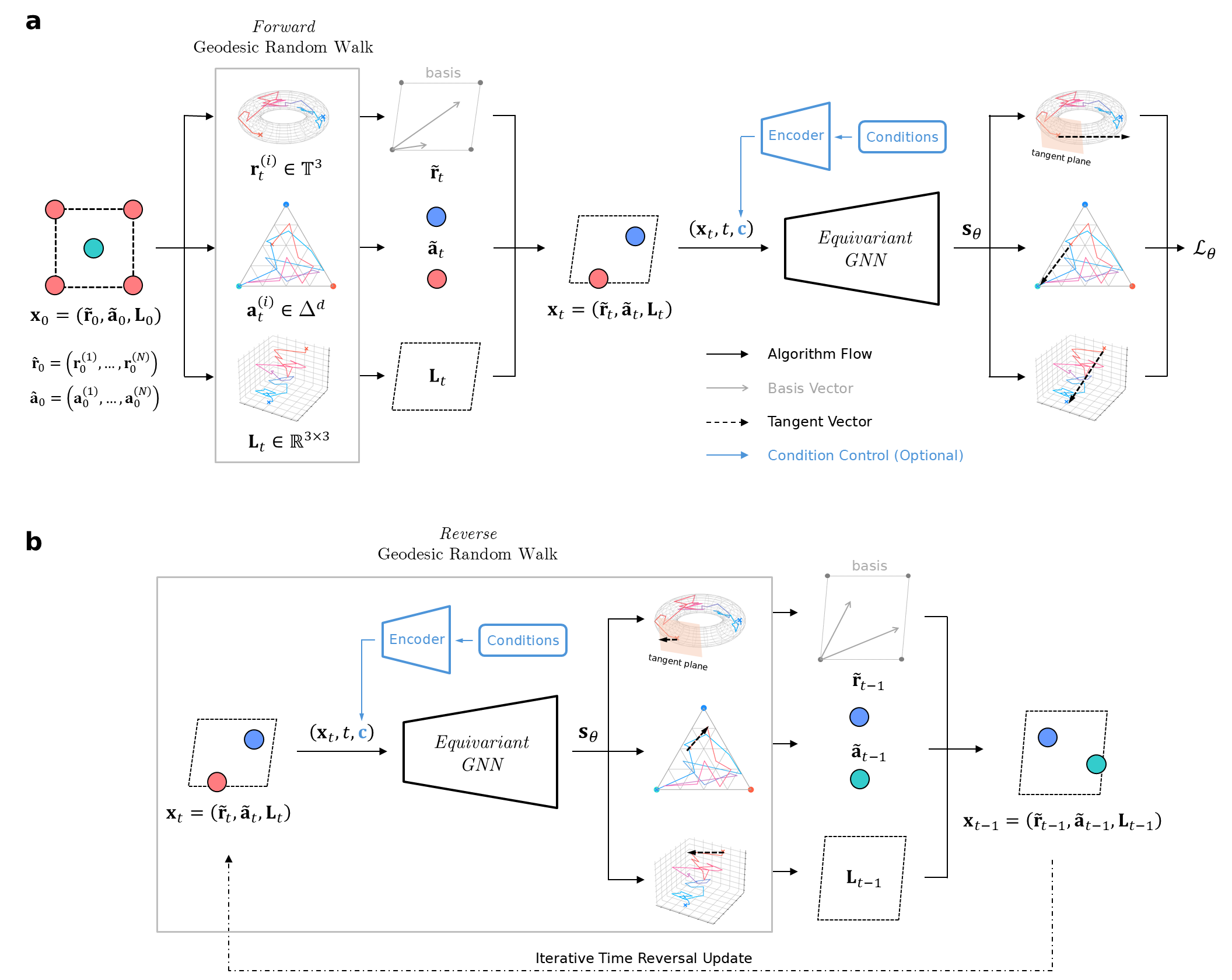}
\caption{
\textbf{Overview of the CrystalGRW workflow.}  
\textbf{(a)} \textbf{Training scheme.} Starting from pristine crystal properties \(\mathbf{x}_0\), which include fractional coordinates \(\mathbf{\tilde{r}}_0 \in \left(\mathbb{T}^3\right)^N\), atomic types \(\mathbf{\tilde{a}}_0 \in \left(\Delta^d\right)^N\), and a lattice matrix \(\mathbf{L}_0 \in \mathbb{R}^{3 \times 3}\), the model applies noise via \emph{forward} geodesic random walks on their respective manifolds, generating corrupted crystal properties \(\mathbf{x}_t\) at a time step \(t\). Here \(N\) is the number of atoms in a unit cell. Then, an equivariant graph neural network (GNN) processes these corrupted properties to predict manifold-specific scores $\mathbf{s}_\theta$ (tangent vectors on each manifold), which are used to compute the training loss \(\mathcal{L}_\theta\).  
\textbf{(b)} \textbf{Sampling scheme.} Beginning from noisy crystal properties \(\mathbf{x}_t\), the trained GNN  computes the trained scores \(\mathbf{s}_\theta\) associated with these inputs. Guided by these scores, a \emph{reverse} geodesic random walk iteratively denoises \(\mathbf{x}_t\) until reaching \(\mathbf{x}_0\), which is the final crystal structure sample.  Specific crystal-structure conditions \(\mathbf{c}\) can be added to the network during training and used to guide attributes (e.g., crystal symmetry) of the generated structures through a classifier-free guidance scheme. 
The 2D torus, 2-simplex (triangle), and 3D Euclidean space manifolds schematically depict the respective manifolds for fractional coordinates, atomic types, and the lattice matrix, serving only as a conceptual visualization rather than an actual representation of these higher-dimensional domains.
}
    \label{fig:model}
\end{figure*}

\subsection{Generating realistic structures}
We train and evaluate the model on the ALEX-MP-20 dataset, which contains 675,204 crystal structures \cite{Zeni2025mattergen}. Each structure contains up to 20 atoms per unit cell, includes up to 10 different elements per compound, and involves only elements with atomic numbers no greater than 83. The dataset is a union of materials from Alexandria materials database \cite{Alexandria1, Alexandria2} and Materials Project database \cite{Jain2013materialsproject} where all structures have an energy above the convex hull ($E_{\textrm{hull}}$) less than 100 meV/atom. We generated 10$^4$ structures where the number of atoms for each structure is sampled from a distribution the same as in the ALEX-MP-20 training set (see Fig.~\ref{fig:alexmp20_stats}(a)). CrystalGRW starts with random atomic positions, atomic types, and lattice matrices, and then iteratively denoises them through GRW for $K$ steps until they approach candidate crystal structures using Algorithm~\ref{algo:reverse_grw_real}. 

Two criteria are used to evaluate whether generated structures are considered to be valid: 1) all interatomic distances be at least 0.5 {\AA} (i.e., structural validity); 2) the total charge of the atoms be neutral \cite{Xie2022cdvae} (i.e., compositional validity). Although these criteria are relatively loose, as noted in Ref.~\cite{Xie2022cdvae}, they provide a useful sanity check for the model’s outcomes. For the $10^4$ structures generated by CrystalGRW, all are structurally valid, and 90.74\% of generated structures are compositionally valid, surpassing the performance of other models (see Table~\ref{tab:validity}). 

\begin{figure*}[ht] \includegraphics[width=0.95\linewidth]{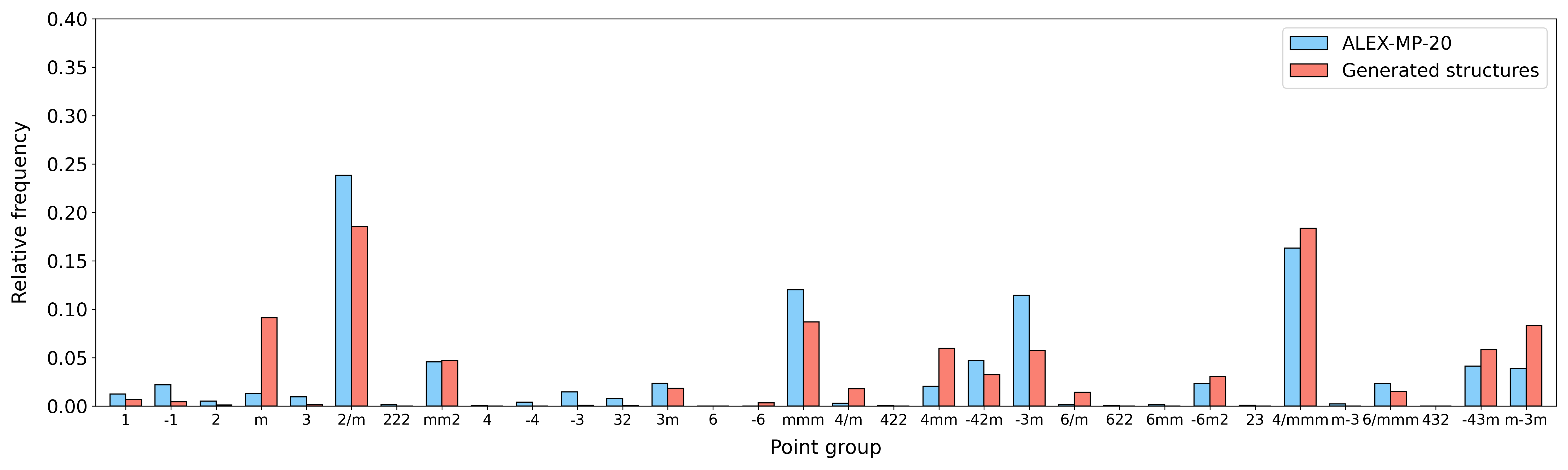}
    \caption{Relative frequency of 32 crystallographic point groups appearing in the ALEX-MP-20 training (blue) and generated structures (pink). The point groups are sorted from low order (left) to high order (right) symmetry.}
    \label{fig:pg_gen_alexmp20}
\end{figure*}

We further identify the symmetry of the generated structures using \texttt{SpacegroupAnalyzer} from \texttt{pymatgen} with symmetry criteria of $\texttt{symprec} = 0.1$ and $\texttt{angle} = 1.0$ \cite{Ong2013pymatgen, Togo2018spglib}. Fig.~\ref{fig:pg_gen_alexmp20} compares the relative frequency of each point group in the ALEX-MP-20 training set and the generated structures. CrystalGRW is able to generate both high-symmetry groups, such as $\bar{6}m2$, $4/mmm$, $6/mmm$, $\bar{4}3m$, and $m\bar{3}m$, and lower-symmetry groups, like $2/m$ and $mm2$, in proportions similar to the ALEX-MP-20 dataset. However, it produces low-symmetry structures excessively, such as $m$, while marginally generating structures in some symmetries, like $\bar{1}$ and groups with a 3-fold principal axis (trigonal lattice). Nonetheless, this result indicates that generated structures possess crystallographic symmetry upon creation.

\begin{table*}[ht]
\caption{Structural and composition validity for 10$^4$ generated structures using the models trained on the MP-20 dataset (and the ALEX-MP-20 dataset for indicated models). RMSD is averaged over 10$^3$ structures that were sampled from the generation set and relaxed using DFT.}
    \begin{ruledtabular}
    \begin{tabular}{ccccc}
        Dataset & Model & Structural (\%) & Composition (\%) & Avg. RMSD ({\AA}) \\
        \hline
        MP-20 & FTCP & 100.00 & 100.00 & 2.42 \\
         & CDVAE & 100.00 & 86.70 & 0.359 \\
        & DP-CDVAE & 99.59 & 85.44 & \\
        & DiffCSP & 100.00 & 83.25 & 0.232 \\
        & FlowMM (500 steps) & 96.86 & 83.24 & \\
        & MatterGen & 100.00 & 84.50 & 0.110 \\
        & CrystalGRW & 100.00 & 85.40 & 0.052  \\
        [0.5em]
        ALEX-MP-20 & DiffCSP & & & 0.104 \\
        & MatterGen & 100.00 & 88.45 & 0.021 \\
        & CrystalGRW & 100.00 & 90.74 & 0.0053 \\
    \end{tabular}
    \end{ruledtabular}
    \label{tab:validity}
\end{table*}

\subsection{Generating stable, unique, and novel structures}

To present the performance of the generative models for crystalline materials, the generated structures should be stable, unique, and novel (S.U.N.) \cite{Zeni2025mattergen}. Firstly, in thermodynamics, materials evolve to the global minimum on the free energy landscape; however, they could be metastable by falling into local minima instead. In terms of energy, materials under certain temperature and pressure are thermodynamically stable if they lie on the convex hull. Nevertheless, the stability of materials can be preliminary considered by $E_{\mathrm{hull}}$. In this work, materials are regarded as stable if their $E_{\mathrm{hull}}$ values are below 100 meV/atom after the DFT relaxations at 0 K and 0 GPa \cite{Zeni2025mattergen}. It is worth noting that this is a relatively loose criterion; in practice, the synthesizability of materials must also account for dynamical and mechanical stability \cite{Mouhat2014bornstability}. Secondly, uniqueness refers to a structure being distinct from other structures in the generation set.
Lastly, the novelty is an evaluation of the number of generated structures that do not match the training set (see Appendix~\ref{sec:unique_novel_appendix}). 

We evaluate novelty by comparing the structures in the generation set with the ALEX-MP dataset and uniqueness by comparing the structures within the generation set (details in Appendix~\ref{sec:unique_novel_appendix}). Fig.~\ref{fig:novelunique_mp20}(a) shows that approximately 48\% of the generated structures (from $10^3$ to $10^4$ samples) are novel. Additionally, around 90\% of $10^3$ structures are unique, with this percentage decreasing to 76\% for $10^4$ samples. Consequently, the percentage of generated structures that are both unique and novel is 44.6\% for $10^3$ samples, which drops to 36.9\% for $10^4$ samples. Although 107.7K disordered structures from ICSD database are not used for novelty evaluation, the unique and novel performance is lower than MatterGen which is 61\% for 10$^4$ generated structures. This may be because CrystalGRW tends to generate highly symmetric structures, as illustrated in Fig.~\ref{fig:pg_gen_alexmp20}. Such symmetry can lead to degeneracy, potentially reducing the uniqueness of the generated structures.

Moreover, we compute the $E_{\mathrm{hull}}$ for 1436 structures using DFT calculation (see Section~\ref{sec:dft}).
Fig.~\ref{fig:novelunique_mp20}(b) illustrates a distribution of the $E_{\mathrm{hull}}$ where 95.0\% of the relaxed structures have $E_{\mathrm{hull}}$ lower than 100 meV/atom. We find that the S.U.N. rate is 37.72\%. In contrast, MatterGen reported 38.57\% for the S.U.N. rate, where novelty was determined by comparing the generated structures with the combined Alex-MP-ICSD dataset, which includes 107.7K disordered structures requiring a license for access \cite{Zeni2025mattergen}. Although our uniqueness and novelty are diminished by symmetrical degeneracy, most of our generated structures are stable, acheiving a S.U.N. rate comparable to MatterGen's. We present examples of S.U.N. structures--metal alloys, cabides, oxides, and hydrides--in Table~\ref{tab:novel_structures} and Fig.~\ref{fig:novel_structures}.

\begin{figure}[ht] \includegraphics[width=0.95\linewidth]{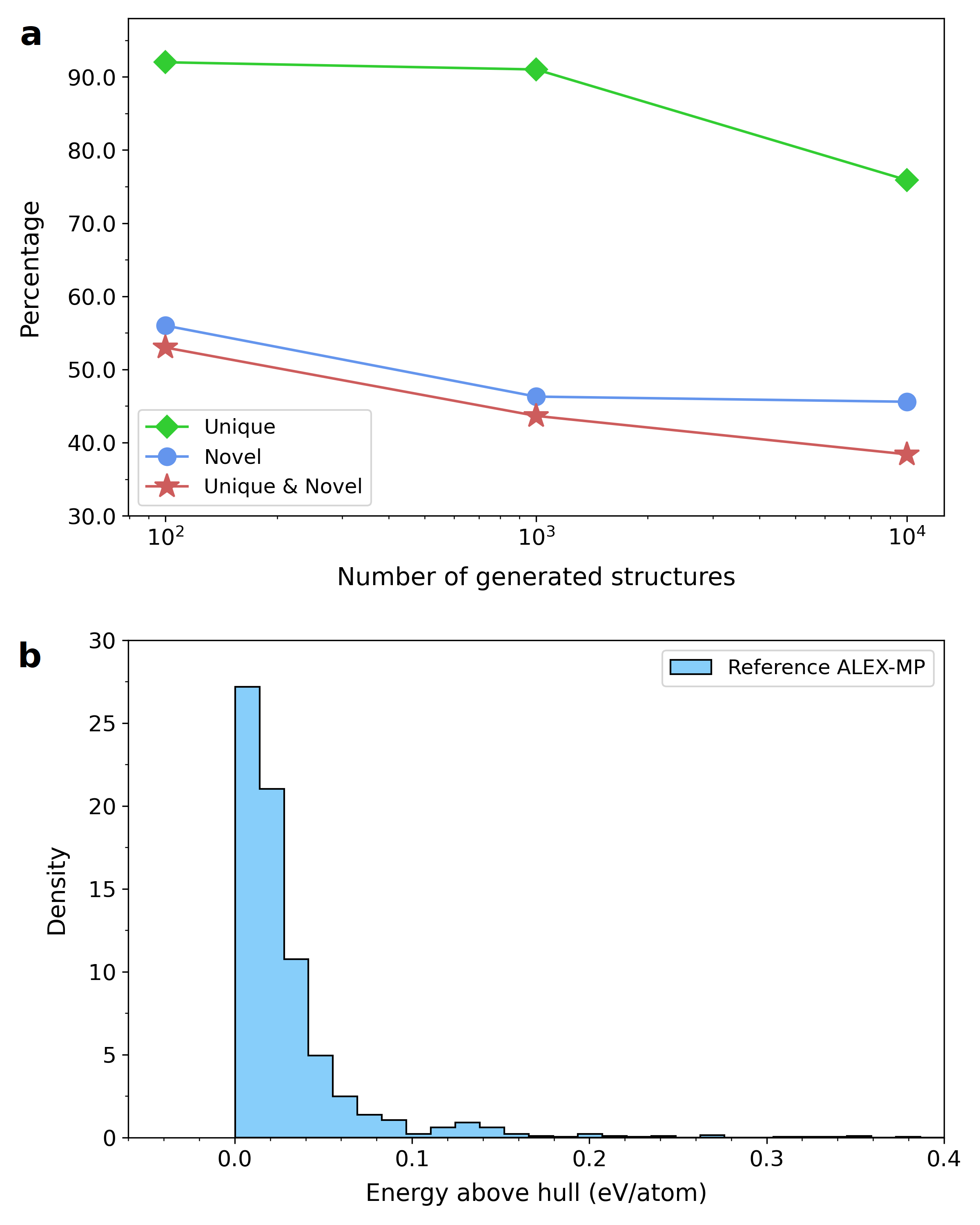}
    \caption{(a) Percentage of generated structures that areunique (green),  novel (blue), and novel and unique (red) with respect to the number of samples. (b) Distribution of $E_{\mathrm{hull}}$ for unique and novel structures computed using the GGA/GGA+$U$ mixing scheme. $E_{\mathrm{hull}}$ is referenced with the convex hull of materials in ALEX-MP dataset.}
    \label{fig:novelunique_mp20}
\end{figure}

\subsection{Generating DFT ground state structures}
\begin{figure}[ht]
 \includegraphics[width=0.95\linewidth]{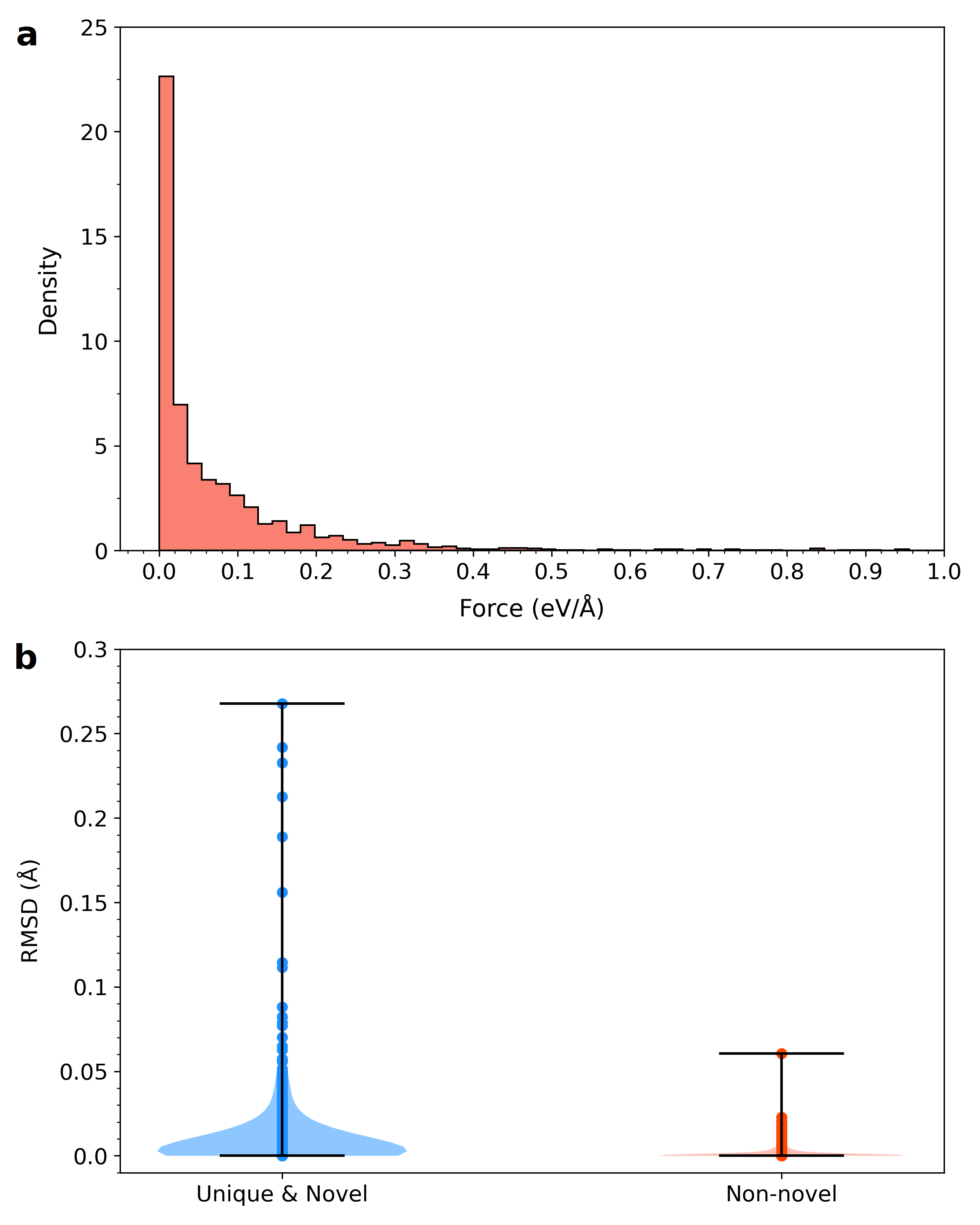}
    \caption{(a) Distribution of DFT force magnitudes (in eV/Å), averaged over all atoms for each generated structure. (b) RMSD between generated and DFT-relaxed structures.}
    \label{fig:force_rmsd}
\end{figure}

One of the primary goals of generative models is to generate crystal structures that are close to DFT ground states. Ground state structures are in equilibrium under thermodynamic conditions where the total force acting on every atom should be 0 eV/{\AA}. From the DFT calculation, we extracted the forces acting on atoms in the generated structures and averaged the magnitude of the force over all atoms. As shown in Fig.~\ref{fig:force_rmsd}(a), 22.94\% and 42.60\% of the samples have forces $\leq 0.002$ and $\leq 0.02$ eV/{\AA}, respectively. DFT relaxation is typically considered converged when the norm of all the forces acting on every atom in a unit cell is $\leq 0.02$ eV/{\AA} (see Section~\ref{sec:dft}), indicating that nearly half of the structures were already close to their DFT ground states upon generation. Machine learning force fields \cite{Batatia2024mace, Deng2023chgnet, Park2024sevennet, Chen2022megnet} enable fast and accurate force computations, so we also employ them to calculate the forces of the generated structures, as illustrated in Fig.~\ref{fig:mlff_force} whose results exhibit similarly to the DFT-calculated forces.

Another metric to determine the degree of structural similarity between generated and DFT-relaxed structures is the root mean square distance (RMSD), which measures the average atomic displacement between two structures. It can be evaluated using \texttt{StructureMacher} from \texttt{pymatgen}. Table~\ref{tab:validity} presents the average RMSD over $10^3$ structures is 0.052 and 0.0053 {\AA} for the models trained on MP-20 and ALEX-MP-20 datasets, respectively. Fig.~\ref{fig:force_rmsd}(b) illustrates the RMSD of structures generated by the model trained on ALEX-MP-20 dataset, grouped into two categories. For unique and novel structures, the average RMSD is 0.011 {\AA} with a few outliers above 0.1 {\AA}. Table~\ref{tab:high_rmsd_structures} lists structures with RMSD $>0.1$ {\AA}, most of which are oxides with 14-20 atoms per unit cell and low-symmetry space group, i.e., $P1$, $Cm$, and $C2/m$.  For non-novel structures, the average RMSDs is 0.0014 {\AA}. This demonstrates that CrystalGRW can generate both non-novel and novel structures that closely match their DFT relaxations, outperforming other models in this task.

\subsection{Generating structures with targeted properties}
One advantage of the generative models is their ability to perform inverse design, generating structures with specific targeted properties. To achieve this, we implement the classifier-free guidance method \cite{Ho2021cfg}, which enhances our model to sample crystal structures on given conditions. We demonstrate our model to control 32 crystallographic point groups of the output structures. However, the number of point groups in the dataset are unbalanced, so we create 7 labels, consisting of $n_1$, $n_2$, $n_3$ for three rotational axes, $m_h$, $m_v$, $m_d$ for horizontal, vertical, and diagonal mirror planes, respectively, and $i$ for an inversion, to represent each point group \cite{DeGraef2012_crystallography}. Figs.~\ref{fig:alexmp20_stats}(b-c) shows that the training set is more balanced when labeled by these 7 labels. (Details on the implementation of conditional control within the graph neural network are provided in section~\ref{sec:pg_control}.)

\begin{figure}[h]
\includegraphics[width=0.9\linewidth]{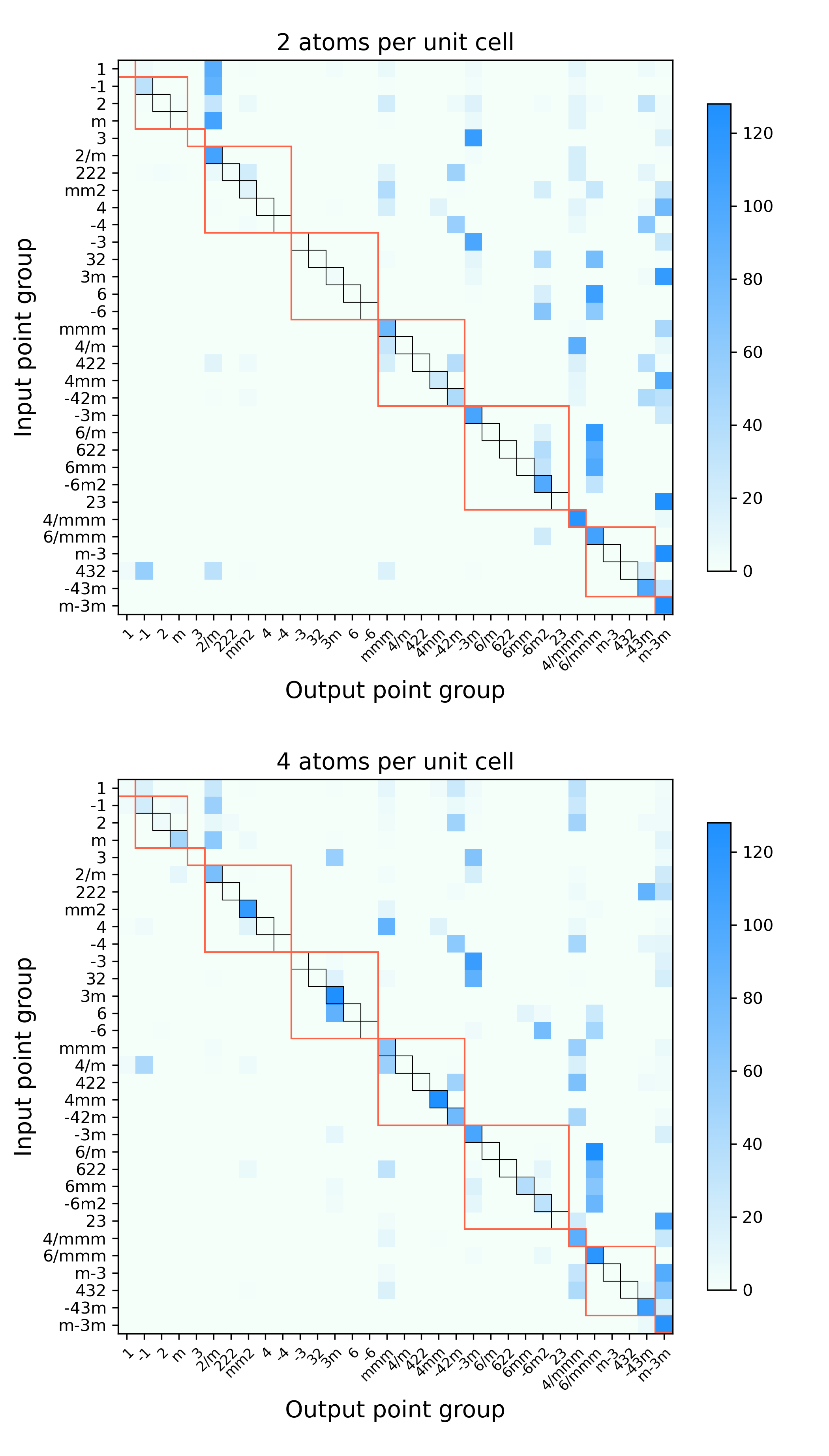}
    \caption{Heat maps illustrate the input conditions, represented as the guiding point groups of output structures (vertical axis), versus the actual point groups of the generated structures (horizontal axis). The generated structures are constrained to contain 2 atoms per unit cell (top) and 4 atoms per unit cell (bottom). Black diagonal blocks indicate cases where the input and output point groups match, while orange diagonal blocks highlight instances where the input and output point groups have the same symmetry order. The color bar denotes the number of structures, ranging from 0 to 128.}
    \label{fig:pg_alexmp20_gs05}
\end{figure}

We generate 128 structures for each point group, specifying 2 and 4 atoms per unit cell and guidance strength $w=0.5$ \cite{Ho2021cfg}. Each row of the heat maps in Fig.~\ref{fig:pg_alexmp20_gs05} presents the output point groups of the generated structures guided (conditioned) by the input point groups. Notably, entries below the diagonal indicate cases where the output symmetry is lower than the input symmetry, while entries above the diagonal indicate higher output symmetry relative to the input.  For both 2 and 4 atoms per unit cell, the generated structures match the input point groups for frequently occurring symmetries in the dataset, such as $m\bar{3}m$, $\bar{4}3m$, $6/mmm$, $\bar{3}m$, $mmm$, and $2/m$ (as shown in Fig.~\ref{fig:pg_gen_alexmp20}). Additionally, $mm2$ and $3m$ also well match for 4-atom structures. However, for less common point groups such as $m\bar{3}$, $\bar{6}$, $\bar{4}$, $\bar{3}$, and $\bar{1}$, the generated structures appear to have higher symmetry that share the same pricipal rotation axis such as $m\bar{3}m$, $6/mmm$, $\bar{4}2m$, $\bar{3}m$, and $2/m$, respectively. 

Table~\ref{tab:pg_percent_match} reports success rates (recalls) of the output structures whose point group symmetry matches the input point group, alongside the percentage of occurrence in the training set for 2 and 4 atoms per unit cell. The results show that point groups with low success rates, such as 4, 23, and $m\bar{3}$, rarely appear or are absent from the training set, whereas point groups with high success rates tend to be those that appear more frequently. Micro F1 scores for each point group are also computed, where true positives are defined as matches between generated and input point groups; false negatives occur when the point group does not match the input; and false positives occur when structures controlled for other point groups appear as the considered point group. Therefore, the $m\bar{3}m$ point group yields a low F1 score despite its high success rate, as structures tend to be $m\bar{3}m$ even they are controlled to be other point groups (especially from absent point groups). On average, the macro F1 scores are 0.130 and 0.216 for 2- and 4-atom unit cells, respectively. The weighted F1 scores are 0.343 and 0.376 for 2- and 4-atom unit cells, respectively, where the weights correspond to the ratios of each point group's occurrence in the training set. Tables~\ref{tab:pg_2atoms} and \ref{tab:pg_4atoms} present examples of structures with 2 and 4 atoms per unit cell, respectively, where the output space group reflects the symmetry imposed by the input point group.

\section{Conclusions}
In summary, we introduced CrystalGRW, a generative model leveraging diffusion on Riemannian manifolds and EquiformerV2 to produce crystal structures with desired properties. CrystalGRW generates unique and novel structures, most of which have $E_{\textrm{hull}}$ below 100 meV. These structures are also close to DFT ground states, outperforming other generative models. We also demonstrated that the properties of generated structures can be controlled by the input conditions, illustrated with the point groups, highlighting the potential to accelerate materials discovery by enabling inverse design through the property guidance.

\section{Methods}
\subsection{Riemannian score-based generative model}
\label{sec:RSGM}

We consider data \(\mathbf{x}_0 \in \mathcal{M}\), where \(\mathcal{M}\) is a Riemannian manifold that may be compact, such as a torus (for fractional coordinates) or a finite-volume hypercube (for atomic types), or non-compact Euclidean space (for lattice matrices). Our goal is to learn a generative model for \(\mathbf{x}_0\) through a score-based diffusion process defined on \(\mathcal{M}\). In the continuous-time limit, the forward noising mechanism can be viewed as an SDE on \(\mathcal{M}\), whose reverse-time counterpart depends on the score \(\nabla_{\mathbf{x}_t}\log p_t(\mathbf{x}_t)\). In particular, one performs a forward (\emph{noising}) diffusion phase that starts with real data \(\mathbf{x}_0\) and evolves it over \([\eta,T]\) until it resembles a tractable noise distribution at \(t=T\) where $\eta \approx 0$. Then, one applies a reverse (\emph{sampling}) diffusion phase, beginning from that noise distribution at \(t=T\) and integrating backward to yield new samples at \(t=0\).  

To implement these phases in discrete time, we employ a \emph{geodesic random walk} (GRW) that updates \(\mathbf{x}_k\) along geodesics in \(\mathcal{M}\), adding drift and noise within the tangent space. 

\vspace{0.5em}
\noindent {\bf Forward geodesic random walk (noising)} Let \(\gamma = T/K\) be the step size, with \(k=0,1,\ldots,K{-}1\) so that \(t = k\gamma\). The forward update is
\begin{equation}
\label{eq:forward_grw}
\begin{aligned}
   \mathbf{W}_{k+1}^{\mathcal{M}} &= \gamma\,\mathbf{b}^{\mathcal{M}}\!\bigl(\mathbf{x}_k,t\bigr)\;+\;\sqrt{\gamma}\,\sigma^{\mathcal{M}}(t)\,\bm{\epsilon}_k,\\
   \mathbf{x}_{k+1} &= \exp_{\mathbf{x}_k}^{\mathcal{M}}\bigl(\mathbf{W}_{k+1}^{\mathcal{M}}\bigr),
\end{aligned}
\end{equation}
where \(\bm{\epsilon}_k \sim \mathcal{N}(\mathbf{0},\,\mathbf{I})\) is an i.i.d.\ Gaussian in the tangent space \(T_{\mathbf{x}_k}\mathcal{M}\), and the noise schedule \(\sigma^{\mathcal{M}}(t)\) specifies how the noise amplitude evolves with time (details in Appendix \ref{sec:appendix_sde}). The exponential map \(\exp_{\mathbf{x}_k}^{\mathcal{M}}\) displaces \(\mathbf{x}_k\) by the tangent vector \(\mathbf{W}_{k+1}^{\mathcal{M}}\).  A crucial choice is the manifold-dependent drift \(\mathbf{b}^{\mathcal{M}}(\mathbf{x},t)\). 

On {\it compact manifolds}, such as tori or closed hypercubes, we typically set \(\mathbf{b}^{\mathcal{M}} = \mathbf{0}\), which is sufficient to make the random walk  become ergodic yielding the late-time noise distribution that approaches the uniform distribution. 

On {\it non-compact} manifolds, the late-time density will converge to a Riemannian normal distribution \cite{Pennec2006} by choosing  $\mathbf{b}^{\mathcal{M}} = -\frac{1}{2}\nabla_{\mathbf{x}}U(\mathbf{x})$ with the potential $U(\mathbf{x}) \propto d_{\mathcal{M}}(\mathbf{x},\boldsymbol{\mu})^2$, where $d_{\mathcal{M}}(\mathbf{x},\boldsymbol{\mu})$ is the geodesic distance and $\boldsymbol{\mu}\in\mathcal{M}$ is a mean location. The forward noising process then follows overdamped Langevin dynamics that eventually converges to the stationary density that is proportional to $ e^{-\,U(\mathbf{x})}$. This Riemannian normal distribution reduces to the standard Gaussian distribution on $\mathbb{R}^d$ when the geodesic distance becomes the usual Euclidean norm. In practice, we further simplify by ignoring $\nabla_{\mathbf{x}}U(\mathbf{x})$ and setting $\mathbf{b}^{\mathbb{R}^d}=0$ directly (as in a variance-exploding approach \cite{Song2021}) to ensure a wide final distribution. For this no-drift approximation, see Appendix~\ref{sec:adaptive_t}b.

After \(K\) iterations of Eq.~\eqref{eq:forward_grw}, the state \(\mathbf{x}_K\) approximates a tractable noise distribution: uniform on compact manifolds, or Gaussian-like on non-compact ones. 

\vspace{0.5em}
\noindent {\bf Reverse geodesic random walk (sampling)} According to Föllmer approach \cite{Follmer1985, Cattiaux2023}, reversing the forward process yields
\begin{equation}
\label{eq:reverse_step}
\begin{aligned}
  \mathbf{W}_{k}^{\mathcal{M}} &= \gamma \Bigl[-\,\mathbf{b}^{\mathcal{M}}\bigl(\mathbf{x}_{k+1},\,T{-}k\gamma\bigr) \;\\ 
  &\quad\;+\;\,\nabla_{\mathbf{x}}\!\log p_{T{-}k\gamma}\bigl(\mathbf{x}_{k+1}\bigr)\Bigr]\\
  &\quad+\;\sqrt{\gamma}\,\sigma^{\mathcal{M}}(T{-}k\gamma)\,\bm{\epsilon}_k,\\
  \mathbf{x}_{k} &= \exp_{\mathbf{x}_{k+1}}^{\mathcal{M}}\bigl(\mathbf{W}_{k}^{\mathcal{M}}\bigr),
\end{aligned}
\end{equation}
where \(\nabla_{\mathbf{x}}\log p_{t}(\mathbf{x})\) is the \emph{score} of the intermediate distribution \(p_{t}\). Note that the sign of \(\mathbf{b}^{\mathcal{M}}\) switches in reverse time.  We next discuss how to approximate \(\nabla_{\mathbf{x}}\log p_{t}\) with a learnable score function of each manifold \(\mathbf{s}^{\mathcal{M}}_{\theta}(\mathbf{x},t)\approx\nabla_{\mathbf{x}}\!\log p_{t}(\mathbf{x})\). 

\vspace{0.5em}
\noindent{\bf Training via Riemannian score matching}
To train \(\mathbf{s}^{\mathcal{M}}_{\theta}\), we adopt a Riemannian score-matching approach, such as Varadhan’s asymptotic \cite{Bortoli2022rsgm, Varadhan1967}, which gives a manifold-specific loss function 
\begin{equation}
\label{eq:rsgm_loss}
\mathcal{L}^{\mathcal{M}}_\theta
=\;
\mathbb{E}\bigl[
  \bigl\|
    \mathbf{s}^{\mathcal{M}}_{\theta}(\mathbf{x}_t,t)
    \;-\;
    \tfrac{1}{t}\,\exp_{\mathbf{x}_t}^{-1,\mathcal{M}}\bigl(\mathbf{x}_0\bigr)
  \bigr\|^{2}
\bigr].
\end{equation}
Here, one uniformly samples a random time \(t\in[\eta,T]\), applies Eq.~\eqref{eq:forward_grw} up to that time $t$, and compares \(\mathbf{s}^{\mathcal{M}}_{\theta}\bigl(\mathbf{x}_t,t\bigr)\) to 
\begin{equation}
    \tfrac{1}{t}\,\mathbf{v}^{\mathcal{M}}_t \equiv \tfrac{1}{t}\,\exp_{\mathbf{x}_t}^{-1,\mathcal{M}}\bigl(\mathbf{x}_0\bigr), 
\end{equation} the latter being the manifold-specific inverse exponential map from \(\mathbf{x}_t\) to \(\mathbf{x}_0\), whose analytical expressions for relevant manifolds are given in Appendix~\ref{appendix:manifolds}. 

We set $\eta=10^{-6}$ rather than $0$ for three key reasons. 
First, it prevents division by zero in the term $\tfrac{1}{t}$ and help avoid pathologically large gradients arising from Varadhan’s asymptotic near $t=0$. 
Second, it ensures a nontrivial denoising task at the earliest time $t=\eta$ (instead of an identity mapping at $t=0$). And lastly, it promotes stable training signals and faster convergence, as the model never sees fully uncorrupted data in its noised phase.

\vspace{0.5em}
\noindent {\bf Variation on adaptive timesteps}  The RSGM algorithm \cite{Bortoli2022rsgm} suggests that each diffusion step can be split into several smaller random-walk increments to enhance numerical stability. 
However, in our experiments, we find that training the model by performing only a single (final) random-walk update per diffusion step is both more computationally efficient and converges faster. Moreover, in reverse diffusion, generated crystal structures become more realistic by using an adaptive timestep ($\gamma \propto k^{\xi}$) for sampling like in previous studies such as in \cite{DBLP:journals/corr/abs-2201-06503} and \cite{DBLP:journals/corr/abs-2106-03802}. See the details in Algorithm \ref{algo:reverse_grw} and Appendix \ref{sec:adaptive_t}.

\subsection{Implementation in CrystalGRW}

In our \emph{CrystalGRW} implementation, we let \(\mathbf{x} = (\mathbf{\tilde r}, \mathbf{\tilde A}, \mathbf{L})\) reside on three distinct manifolds \(\bigl((\mathbb{T}^3)^N,\;(\mathbb{C}^d)^N,\;\mathbb{R}^{3\times 3}\bigr)\), reflecting the natural domains of fractional coordinates, atomic types, and lattice parameters, respectively. We define a manifold-specific drift \(\mathbf{b}^{\mathcal{M}}(\mathbf{x},t)\) that vanishes on the compact manifolds but equals \(-\frac{1}{2}\nabla U\) for the Euclidean one, and train an equivariant graph neural network \(\mathbf{s}_{\theta}\) (EquiformerV2 \cite{Liao2024equiformerv2}) which provides separate outputs for each manifold slice. To learn the score function jointly across all manifolds, we minimize the total loss
\[
   \mathcal{L}_\theta 
   \;=\;
   \lambda_{(\mathbb{T}^3)^N}\mathcal{L}_\theta^{(\mathbb{T}^3)^N} 
   \;+\;
   \lambda_{(\mathbb{C}^d)^N}\mathcal{L}_\theta^{(\mathbb{C}^d)^N} 
   \;+\;
   \lambda_{\mathbb{R}^{3 \times 3}}\mathcal{L}_\theta^{\mathbb{R}^{3\times 3}},
\]
where each term follows the Riemannian score-matching objective in Eq.~\eqref{eq:rsgm_loss} and each $\lambda_{\mathcal{M}}$ is a hyperparameter that weights the contribution of the loss on each manifold. However, for the finite hypercube manifold describing atomic types, we find it both more efficient and more accurate to train an estimator network in place of a direct score predictor; details of this atomic-type denoising approach appear in Appendix~\ref{sec:appendix_atomic_types}. Appendix~\ref{appendix:equiformerv2} details the equivariant network architecture, and Algorithm~\ref{algo:train} provides the step-by-step training procedure.

After completing the training procedure, we sample from the trained generative model by first drawing a final noisy state $\mathbf{x}_K$ from either a uniform prior (if the manifold is compact) or a Gaussian prior (if it is Euclidean). We then run the reverse geodesic random walk in Eq.~\eqref{eq:reverse_step}, using the learned score $\mathbf{s}_{\theta}$, to generate a new sample $\mathbf{x}_0$. The complete procedure is detailed in Algorithm~\ref{algo:reverse_grw}.

\vspace{0.5em}
\noindent\textbf{Algorithm modifications}
A key challenge lies in generating atomic types. As discussed in Appendix~\ref{sec:appendix_atomic_types}, our experiments revealed that using Algorithms~\ref{algo:train} and \ref{algo:reverse_grw} to generate atomic types, particularly for perovskite compositions, resulted in suboptimal performance. To address this, we adapted the process by training atomic types against their ground truths, as outlined in Algorithm~\ref{algo:train_real}, and sampling them using Algorithm~\ref{algo:reverse_grw_real}. Detailed discussions and methodologies are provided in Appendix~\ref{sec:appendix_atomic_types}. All reported results in Figs.~\ref{fig:pg_gen_alexmp20}-\ref{fig:force_rmsd} are based on the modified Algorithms~\ref{algo:train_real} and \ref{algo:reverse_grw_real}.

\vspace{0.5em}
\noindent\textbf{Condition guidance}
When conditional sampling is desired (e.g., for controlling point groups or compositions), we apply \emph{classifier-free guidance} \cite{Ho2021cfg} in our manifold-aware pipeline. Let $\mathbf{s}_{\theta}(\mathbf{x}\mid\mathbf{c})$ be the score for a condition $\mathbf{c}$, and let $\mathbf{s}_{\theta}(\mathbf{x}\mid\varnothing)$ be the unconditional score (i.e., with a null condition). We construct a \emph{condition-guided} score by combining a conditional prediction $\mathbf{s}_{\theta}(\mathbf{x}\mid\mathbf{c})$ with an unconditional one $\mathbf{s}_{\theta}(\mathbf{x}\mid\varnothing)$ as
\begin{equation} 
\label{eq:conditional_score}
\widetilde{\mathbf{s}}_{\theta}\bigl(\mathbf{x}\,\big\vert\,\mathbf{c}\bigr)
    \;=\;
    (1 + w)\,\mathbf{s}_{\theta}(\mathbf{x}\mid\mathbf{c})
    \;-\;
    w\,\mathbf{s}_{\theta}(\mathbf{x}\mid\varnothing),
\end{equation}
where $w$ is the guidance strength. During training, we randomly replace a fraction (e.g.\ 10\%) of the conditional samples with a null condition $\varnothing$, obtained by zeroing out the condition vector. Thus, if the batch size is 10, about 1 sample per batch is trained with $\varnothing$ while the others include the true condition $\mathbf{c}$. In sampling process, we interpolate between $\mathbf{s}_{\theta}(\mathbf{x}\mid\mathbf{c})$ and $\mathbf{s}_{\theta}(\mathbf{x}\mid\varnothing)$ according to $w$ using Eq.~\eqref{eq:conditional_score}, thereby controlling how strongly the generation is influenced by $\mathbf{c}$. This procedure naturally extends to our manifold setting by applying $\widetilde{\mathbf{s}}_{\theta}$ in the reverse diffusion step Eq.~\eqref{eq:reverse_step}, and ensures consistent conditional guidance across all manifold components.

\begin{algorithm}[!h]
\noindent\hrulefill\par\nobreak\vskip-10pt
\caption{\centering Forward Geodesic Random Walk (\emph{Noising}) for Multi-Manifold States}
\vskip-7pt\hrulefill
\label{algo:forward_grw}
\begin{algorithmic}[1]
\REQUIRE \Statex \begin{itemize}
    \item $\mathbf{X}_0 =\bigl(\mathbf{\tilde{r}}_0,\;\mathbf{\tilde{A}}_0,\;\mathbf{L}_0\bigr)$, each on manifolds $\left(\mathbb{T}^3\right)^N, \left(\mathbb{C}^d\right)^N, \mathbb{R}^{3\times3}$ respectively
    \item  Total steps $K$, final time $T$ (\textit{hence} $\gamma = T/K$)
\end{itemize}
\vspace{5pt}
\Statex
\FOR{$k = 0$ to $K-1$}
  \State $t = k\,\gamma$
  \State Get drift coefficients
        \[
        \bigl(\underbrace{\mathbf{0}}_{\left(\mathbb{T}^3\right)^N},
            \underbrace{\mathbf{0}}_{\left(\mathbb{C}^d\right)^N},
            \underbrace{\mathbf{b}^{\mathbb{R}^{3\times 3}}}_{\approx \mathbf{0}\in\mathbb{R}^{3\times3}}\bigr)
        \gets \mathbf{b}\bigl(\mathbf{X}_{k}, t\bigr)
       \]
\Rulecomment{See Appendix \ref{sec:adaptive_t}b for no drift approximation $\quad \quad \quad$}
  \State Apply to all manifolds
  \[
  \begin{aligned}
  \mathbf{W}^{\mathcal{M}}_{k+1} = \gamma\mathbf{b}^{\mathcal{M}} + \sqrt{\gamma}\sigma^{\mathcal{M}}(t)\bm{\epsilon}_k, \;\;\; \bm{\epsilon}_k \sim \mathcal{N}\bigl(\mathbf{0},\mathbf{I}\bigr)
  \end{aligned}
  \]
  \Rulecomment{Specific parameterizations of $\sigma^{\mathcal{M}}(t)$ in Appendix \ref{sec:appendix_sde}}
  \State $(\mathbf{\tilde r}_k,\,\mathbf{\tilde A}_k,\,\mathbf{L}_k) \gets \mathbf{X}_{k}$
  \State Apply manifold-specific exponential maps
  \[
  \begin{aligned}
    \mathbf{r}_{k+1}^{(i)} &= \exp_{\mathbf{r}_{k}^{(i)}}^{\mathbb{T}^3}\!\bigl(\mathbf{W}_{k+1}^{\mathbb{T}^3, (i)}\bigr), \\
    \mathbf{A}_{k+1}^{(i)} &= \exp_{\mathbf{A}_{k}^{(i)}}^{\mathbb{C}^d}\!\bigl(\mathbf{W}_{k+1}^{\mathbb{C}^d, (i)}\bigr), \\
    \mathbf{L}_{k+1} &= \exp_{\mathbf{L}_{k}}^{\mathbb{R}^{3\times 3}}\!\bigl(\mathbf{W}_{k+1}^{\mathbb{R}^{3\times 3}}\bigr)
  \end{aligned}
  \]
  
  \State $\mathbf{X}_{k+1} = (\mathbf{\tilde r}_{k+1},\;\mathbf{\tilde A}_{k+1},\;\mathbf{L}_{k+1})$
\ENDFOR
\State \textbf{return} $\mathbf{X}_K = (\mathbf{\tilde r}_K,\;\mathbf{\tilde A}_K,\;\mathbf{L}_K)$
\end{algorithmic}
\end{algorithm}

\begin{algorithm}[t]
\noindent\hrulefill\par\nobreak\vskip-10pt
\caption{\centering \emph{Training} Multi-Manifold Crystal Properties}
\vskip-7pt\hrulefill
\label{algo:train}
\begin{algorithmic}[1]
\REQUIRE \Statex 
\begin{itemize}
    \item $\mathbf{x}_0 =\bigl(\mathbf{\tilde{r}}_0,\;\mathbf{\tilde{a}}_0,\;\mathbf{L}_0\bigr)$, each on manifolds $\left(\mathbb{T}^3\right)^N, \left(\Delta^d\right)^N, \mathbb{R}^{3\times3}$ respectively

    \item Total steps $K$
\end{itemize}
\vspace{5pt}   
\State $\mathbf{\tilde{A}}_0 = \mathcal{H}(\mathbf{\tilde{a}}_0)$
\State $\mathbf{X}_0 =\bigl(\mathbf{\tilde{r}}_0,\;\mathbf{\tilde{A}}_0,\;\mathbf{L}_0\bigr)$

\REPEAT
	    \State \(t \sim  \mathrm{Uniform}(\eta, T)\) 
	    \State \(\bigl(\mathbf{\tilde r}_t,\;\mathbf{\tilde A}_t,\;\mathbf{L}_t\bigr) \gets \mathrm{GRW}\left(\mathbf{X}_0, K, t\right)\) \Rulecomment{Algorithm \ref{algo:forward_grw}}
	    \State Compute inverse exponential maps
	    \[
	    \begin{aligned}
	      \mathbf{v}^{\mathbb{T}^3, (i)}_t \;&=\;
	\exp_{\mathbf{r}_t^{(i)}}^{-1,\mathbb{T}^3}\bigl(\mathbf{r}_0^{(i)}\bigr),\\
	\mathbf{v}^{\mathbb{C}^d, (i)}_t \;&=\;
	\exp_{\mathbf{A}_t^{(i)}}^{-1,\mathbb{C}^d}\bigl(\mathbf{A}_0^{(i)}\bigr),\\
	\mathbf{v}^{\mathbb{R}^{3\times 3}}_t \;&=\;
	\exp_{\mathbf{L}_t}^{-1,\mathbb{R}^{3\times 3}}\bigl(\mathbf{L}_0\bigr)
	    \end{aligned}
	  \]
	\State $\mathbf{\tilde{a}}_t = \mathcal{H}^{-1}(\mathbf{\tilde{A}}_t)$
	\State $\mathbf{x}_t =\bigl(\mathbf{\tilde{r}}_t,\;\mathbf{\tilde{a}}_t,\;\mathbf{L}_t\bigr)$
	\State $\bigl(\mathbf{s}^{\mathbb{T}^3}_{\theta}, \mathbf{s}^{\mathbb{C}^d}_{\theta}, \mathbf{s}^{\mathbb{R}^{3\times 3}}_{\theta}\bigr) \gets \mathbf{s}_{\theta}(\mathbf{x}_t,t)$
	
		\State Compute the loss  \Rulecomment{Varadhan's asymptotic}
		    \[
		    \begin{aligned}
		      \mathcal{L}_\theta
		      &\;=\;
		      \frac{\lambda_{(\mathbb{T}^3)^N}}{N} \sum_{i=1}^N\Bigl\|
		        \mathbf{s}_\theta^{\mathbb{T}^3, (i)}
		        \;-\;
		        \mathbf{v}^{\mathbb{T}^3, (i)}_t/t
		      \Bigr\|^{2}
		      \\
		      &\;+\;
		      \frac{\lambda_{(\mathbb{C}^d)^N}}{N} \sum_{i=1}^N\Bigl\|
		        \mathbf{s}_\theta^{\mathbb{C}^d, (i)}
		        \;-\;
		        \mathbf{v}^{\mathbb{C}^d, (i)}_t/t
		      \Bigr\|^{2}\\
		      &\;+\;
		      \lambda_{\mathbb{R}^{3\times 3}} \Bigl\|
		        \mathbf{s}_\theta^{\mathbb{R}^{3\times 3}}
		        \;-\;
		        \mathbf{v}^{\mathbb{R}^{3\times 3}}_t/t
		      \Bigr\|^{2}
		    \end{aligned}
		  \]

	\State  Take gradient descent step on \( \mathbf{\nabla}_{\theta} \mathcal{L}_{\theta} \)
\UNTIL{converged}
\end{algorithmic}
\end{algorithm}

\begin{algorithm}[h]
\noindent\hrulefill\par\nobreak\vskip-10pt
\caption{\centering Reverse Geodesic Random Walk (\emph{Sampling}) for Multi-Manifold States}
\vskip-7pt\hrulefill
\label{algo:reverse_grw}
\begin{algorithmic}[1]

\REQUIRE \(N, K, T, \xi, \mathbf{s}_{\theta}\)
\vspace{5pt}

\State $\mathbf{\tilde{r}}_T = (\mathbf{r}_T^{(1)}, \dots, \mathbf{r}_T^{(N)}) \sim \mathrm{Uniform}\bigl([0,1)\bigr)^{N \times 3}$ 

\State $\mathbf{\tilde{A}}_T = (\mathbf{A}_T^{(1)}, \dots, \mathbf{A}_T^{(N)})  \sim \text{Uniform}\left([0,1]^d\right)^N$  

\State $\mathbf{L}_T \sim \mathcal{N}\!\Bigl(\mathbf{0},\,\tfrac{h_f + h_0}{2}\,\mathbf{I}\Bigr)$ where $\mathbf{0}, \mathbf{I}\in \mathbb{R}^{3\times 3}$

\State $\mathbf{\tilde a}_{K} = \mathcal{H}^{-1}(\mathbf{\tilde A}_{K})$

  \State $\mathbf{x}_{K} = (\mathbf{\tilde r}_{K},\;\mathbf{\tilde a}_{K},\;\mathbf{L}_{K})$

\FOR{$k = K \text{ to }  1$}
  \State $\gamma = (T/K)k^{\xi}$
  \State $t = \gamma k$
  
  \State $\bigl(\mathbf{s}^{\mathbb{T}^3}_{\theta}, \mathbf{s}^{\mathbb{C}^d}_{\theta}, \mathbf{s}^{\mathbb{R}^{3\times 3}}_{\theta}\bigr) \gets \mathbf{s}_{\theta}(\mathbf{x}_k, t)$
  \State Get drift coefficients
        \[
        \bigl(\underbrace{\mathbf{0}}_{\left(\mathbb{T}^3\right)^N},
            \underbrace{\mathbf{0}}_{\left(\mathbb{C}^d\right)^N},
            \underbrace{\mathbf{b}^{\mathbb{R}^{3\times 3}}}_{\approx \mathbf{0}\in\mathbb{R}^{3\times3}}\bigr)
        \gets \mathbf{b}\bigl(\mathbf{X}_{k}, t\bigr)
       \]

  \State Apply to all manifolds
  \[
  \begin{aligned}
  \mathbf{W}^{\mathcal{M}}_{k-1} = \gamma\Bigl[-\mathbf{b}^{\mathcal{M}} +\mathbf{s}^{\mathcal{M}}_{\theta}
                 \Bigr] + \sqrt{\gamma}\sigma^{\mathcal{M}}(t)\bm{\epsilon}_k, \\ \bm{\epsilon}_k \sim \mathcal{N}\bigl(\mathbf{0},\mathbf{I}\bigr)
  \end{aligned}
  \]
  
  \State $(\mathbf{\tilde r}_k,\,\mathbf{\tilde A}_k,\,\mathbf{L}_k) \gets \mathbf{X}_{k}$
  \State Apply manifold-specific exponential maps
  \[
  \begin{aligned}
    \mathbf{r}_{k-1}^{(i)} &= \exp_{\mathbf{r}_{k}^{(i)}}^{\mathbb{T}^3}\!\bigl(\mathbf{W}_{k-1}^{\mathbb{T}^3, (i)}\bigr), \\
    \mathbf{A}_{k-1}^{(i)} &= \exp_{\mathbf{A}_{k}^{(i)}}^{\mathbb{C}^d}\!\bigl(\mathbf{W}_{k-1}^{\mathbb{C}^d, (i)}\bigr), \\
    \mathbf{L}_{k-1} &= \exp_{\mathbf{L}_{k}}^{\mathbb{R}^{3\times 3}}\!\bigl(\mathbf{W}_{k-1}^{\mathbb{R}^{3\times 3}}\bigr)
  \end{aligned}
  \]
  \State $\mathbf{\tilde a}_{k-1} = \mathcal{H}^{-1}(\mathbf{\tilde A}_{k-1})$

  \State $\mathbf{x}_{k-1} = (\mathbf{\tilde r}_{k-1},\;\mathbf{\tilde a}_{k-1},\;\mathbf{L}_{k-1})$
\ENDFOR
\State \textbf{return} $\mathbf{x}_0 = (\mathbf{\tilde r}_0,\;\mathbf{\tilde a}_0,\;\mathbf{L}_0)$

\end{algorithmic}
\end{algorithm}

\subsection{Crystal graphs}
Here we describe how \(\mathbf{x} = (\mathbf{\tilde r}, \mathbf{\tilde a}, \mathbf{L})\) and \(\mathbf{c}\) are integrated into inputs of graph neural networks. Since GNNs incorporate inductive biases via graph structures \cite{Battaglia2018inductivebiases}, we represent each crystal as a multigraph $\mathcal{G}=(\mathcal{V},\mathcal{E})$. The node set $\mathcal{V}$ includes both atom-centric features and positions, while the edge set $\mathcal{E}$ captures interactions between pairs of atoms. In this multigraph setting, the same pair of nodes can be connected by multiple edges if different crystallographic directions link those atoms. Concretely, we define
\begin{equation}\label{eq:crystal_graphs}
\begin{gathered}
    \mathcal{V} \;=\; \bigl\{(\mathbf{f}^{(i)},\;\mathbf{r'}^{(i)})\;\big|\;\mathbf{f}_i\in\mathbb{R}^N,\;\mathbf{r'}^{(i)}=\mathbf{L}\,\mathbf{r}^{(i)}\in\mathbb{R}^3\bigr\},\\
    \mathcal{E} \;=\; \bigl\{\Delta \mathbf{r'}_{ji}^{(\mathbf{T})}\;\big|\;\Delta \mathbf{r'}_{ji}^{(\mathbf{T})}=\mathbf{r'}^{(j)}-\mathbf{r'}^{(i)}+\mathbf{T};\;\mathbf{r'}^{(j)},\mathbf{r'}^{(i)}\in\mathbb{R}^3\bigr\},
\end{gathered}
\end{equation}
where $\mathbf{f}^{(i)}$ is a node attribute (e.g. atomic number derived from the atomic type $\mathbf{a}^{(i)}$), $\mathbf{r'}^{(i)}$ is the Cartesian position of atom $i$, $\mathbf{T}$ is a translation vector, and $N$ is the number of atoms in a unit cell. The lattice matrix $\mathbf{L}\in\mathbb{R}^{3\times 3}$ maps fractional coordinates $\mathbf{r}^{(i)}$ to $\mathbf{r'}^{(i)}$. Although some previous works represented $\mathbf{L}$ in \(\mathbb{R}^6\) \cite{Zeni2025mattergen, Miller2024flowmm}, we find that diffusing all nine elements of $\mathbf{L}$ yields similar performance. For a detailed discussion of crystal graphs defined by Eq.~\eqref{eq:crystal_graphs}, see Ref. \cite{Pakornchote2023}.

For each node, we concatenate its node attributes (e.g.\ atomic numbers) with a time step $t$ to reflect the diffusion stage. Empirically, appending the corrupted lattice variables to the node attributes further improve model performance. For the conditional control, we similarly append condition vector embeddings $\mathbf{c}$ (e.g.\ point groups) to these node features (see Fig.~\ref{fig:model}), thereby allowing the GNN to handle a condition alongside the atomic attriubutes.

\subsection{DFT setting}
\label{sec:dft}
We evaluated energies and performed structural relaxations using density functional theory (DFT) \cite{hohenberg1964inhomogeneous,kohn1965self}, adopting the projector augmented wave (PAW) method \cite{kresse1999ultrasoft,hobbs2000fully} as implemented in the Vienna \emph{Ab Initio} Simulation Package (VASP) \cite{kresse1996efficiency,kresse1996efficient}. Our calculations followed the \texttt{MPRelaxSet}, a Materials Project workflow \cite{jain2011high}, which includes exchange-correlation functionals and sets a plane-wave energy cutoff of 520\,eV, and energy and force convergence thresholds of $10^{-5}$\,eV and 0.02\,eV/{\AA}, respectively. Brillouin zone sampling was achieved via Monkhorst-Pack $k$-point grids \cite{monkhorst1976special}, maintaining a \emph{k}-point density of 1000\,\emph{k}-points reciprocal per atom for different unit-cell sizes. We used the Perdew--Burke--Ernzerhof (PBE) exchange-correlation functional for the generalized gradient approximation (GGA) \cite{perdew1996generalized} and the GGA with Hubbard \textit{U} corrections (GGA+\textit{U})  \cite{wang2006oxidation,jain2011formation}, where the spin polarization was enabled for the latter method. To evaluate the thermodynamic stability \cite{bartel2022review}, we computed the $E_{\mathrm{hull}}$ using the GGA/GGA+\textit{U} mixing scheme and constructed phase diagrams with \texttt{PhaseDiagram} tool from \texttt{pymatgen} \cite{Ong2008phasediagram, Ong2010phasediagram}. We found that unary compounds such as Mo, Pd, Ir, and Pt have $E_{\textrm{hull}}$ significantly lower than those in the Materials Project database, despite following the MPRelaxSet. Therefore, we excluded structures containing these elements to avoid artifically low $E_{\textrm{hull}}$ in the stability evaluation.

\section*{Data Availability} 
The ALEX-MP dataset for novelty evaluation and the ALEX-MP-20 dataset for training are available at \url{https://github.com/microsoft/mattergen}. The MP-20 dataset for training, including point group labels for conditional control, is available at \url{https://github.com/trachote/crystalgrw}.

\section*{Code Availability} The CryslGRW source code is available at \url{https://github.com/trachote/crystalgrw}.

\bibliography{apssamp}

\begin{acknowledgments}
K.T. and T.B. acknowledge funding support from Thailand Science Research and Innovation Fund Chulalongkorn University (IND\_FF\_68\_285\_2300\_072). T.C. and A.E. acknowledge the funding support from the NSRF via the Program Management Unit for Human Resources \& Institutional Development, Research and Innovation [grant number B39G670016]. A.E. also gratefully acknowledges the supports from the Sci-Super X fund, Faculty of Science, Chulalongkorn University. B.A. acknowledges the financial support from the Swedish Research Council (VR) through Grant No. 2019-05403 and 2023-05194, and from the Swedish Government Strategic Research Area in Materials Science on Functional Materials at Linköping University (Faculty Grant SFOMatLiU No. 2009-00971). 
T.P. and C.S. acknowledge the financial support from the Army Research Office and Air Force Office of Scientific Research (FA9550-23-1-0505). 
The authors acknowledge high performance
computing resources including NVIDIA A100
GPUs from Chula Intelligent and Complex Systems
Lab, Faculty of Science and from the Center for AI in
Medicine (CU-AIM), Faculty of Medicine, Chulalongkorn
University, Thailand. The supporting
computing infrastructure is provided by NSTDA, CU,
CUAASC, NSRF via PMUB [B05F650021, B37G660013]
(Thailand). URL:www.e-science.in.th. We also acknowledge the computing resource from the National Academic Infrastructure for Supercomputing in Sweden (NAISS) partially funded by the Swedish Research Council through grant agreement no. 2022-06725.
This research used resources of the National Energy Research Scientific Computing Center (NERSC), a Department of Energy Office of Science User Facility, using the award GenAI@NERSC.
\end{acknowledgments}

\section*{Author contributions} 
K.T., T.P., and T.C. designed the methodology. K.T., T.P., and N.C. developed the software. K.T., T.P., and C.A. performed the validation and formal analysis. A.E., B.A., C.S., T.B., and T.C. provided resources. T.B. supervised the project. K.T., T.P., and T.C. wrote the original draft. All authors contributed to the review and editing of the manuscript.

\section*{Competing interests} 
The authors declare no competing interests.

\section*{Supplementary information} 
\appendix
\section{Manifolds for Fractional Coordinates, Atomic Types, and Lattice Matrices}
\label{appendix:manifolds}

In this appendix, we describe the manifold structures relevant to our crystal-geometry formulation. Throughout, let $\mathcal{M}$ be a Riemannian manifold equipped with tangent bundles $\mathcal{T}_{\mathbf{x}}\mathcal{M}$ at each point $\mathbf{x}\in\mathcal{M}$. 

\subsection{General definitions}

Suppose $\mathbf{u}\in \mathcal{T}_{\mathbf{x}_1}\mathcal{M}$ is a tangent vector at $\mathbf{x}_1\in\mathcal{M}$. Moving $\mathbf{x}_1$ along $\mathbf{u}$ by a geodesic yields a new point
\begin{equation}
    \mathbf{x}_2 \;=\; \exp^{\mathcal{M}}_{\mathbf{x}_1}(\mathbf{u}),
\end{equation}
where $\exp^{\mathcal{M}}_{\mathbf{x}_1}$ is the exponential map. Conversely, if $\mathbf{x}_2$ is known, we can recover $\mathbf{v}\in \mathcal{T}_{\mathbf{x}_2}\mathcal{M}$ that points back toward $\mathbf{x}_1$ via
\begin{equation}
    \mathbf{v} \;=\; \exp_{\mathbf{x}_2}^{-1, \mathcal{M}}(\mathbf{x}_1).
\end{equation}

Below, we detail how these maps take concrete forms for the Euclidean manifold (for lattice matrices), the $3$-dimensional torus (for fractional coordinates), and a hypercube (for atomic types).

\subsection{Euclidean manifold $\mathbb{R}^{3 \times 3}$ for lattice matrices}

A lattice matrix $\mathbf{L}$ lives in $\mathbb{R}^{3\times3}$. Consider the Euclidean manifold $\mathbb{R}^{d}$. The exponential map from a point $\mathbf{x}_1$ to $\mathbf{x}_2$ along a vector $\mathbf{u}\in\mathbb{R}^d$ simply becomes
\begin{equation}
   \mathbf{x}_2 \;=\; \exp_{\mathbf{x}_1}^{\mathbb{R}^{d}}(\mathbf{u})
   \;=\;\mathbf{x}_1 + \mathbf{u},
\end{equation}
and the corresponding inverse exponential map is
\begin{equation}
    \mathbf{v} \;=\; \exp_{\mathbf{x}_2}^{-1,\mathbb{R}^d}(\mathbf{x}_1)
    \;=\; \mathbf{x}_1 - \mathbf{x}_2.
\end{equation}
Because $\mathbb{R}^{d}$ is flat, geodesics are just straight lines.

\subsection{3D torus $\mathbb{T}^3$ for fractional coordinates}\label{sec:torus}

Fractional coordinates of crystal lattices have an intrinsic periodicity, which we interpret as a 3D torus, $\mathbb{T}^{3}$. In this compact manifold, one often wraps coordinates modulo 1 to enforce periodic boundaries. For an infinitesimal tangent vector $\mathbf{u}$ at $\mathbf{x}_1\in\mathbb{T}^3$, the exponential map can be explicitly written as
\begin{equation}
    \mathbf{x}_2 
    \;=\; \exp_{\mathbf{x}_1}^{\mathbb{T}^{3}}(\mathbf{u})
    \;=\; \cos\bigl(\|\mathbf{u}\|\bigr)\,\mathbf{x}_1 
    + \sin\bigl(\|\mathbf{u}\|\bigr)\,\frac{\mathbf{u}}{\|\mathbf{u}\|},
\end{equation}
with the inverse map
\begin{equation}
    \mathbf{v}
    \;=\; \exp_{\mathbf{x}_2}^{-1,\mathbb{T}^{3}}(\mathbf{x}_1)
    \;=\;\arccos\bigl(\langle \mathbf{x}_2,\mathbf{x}_1\rangle\bigr)\,
    \frac{\mathbf{x}_1 - \langle\mathbf{x}_2,\mathbf{x}_1\rangle\,\mathbf{x}_2}
         {\bigl\|\mathbf{x}_1 - \langle\mathbf{x}_2,\mathbf{x}_1\rangle\,\mathbf{x}_2\bigr\|}.
\end{equation}
In practice, one usually implements these periodic transformations by coordinate wrapping (modulo 1) rather than the analytic trigonometric form, but the latter is useful for conceptualizing geodesics on $\mathbb{T}^3$.

\subsection{$d$-simplex $\Delta^d$ and $d$-hypercube $\mathbb{C}^d$ for atomic types}
\label{sec:hypercube}

Each atomic site $i$ can be occupied by any one of \(d+1\) possible species. One can naturally encodes this uncertainty in a probability vector \(\mathbf{a}^{(i)} = (a_{1}^{(i)}, a_{2}^{(i)}, \dots, a_{d}^{(i)})\), where \(a_{\alpha}^{(i)} \ge 0\) for all \(\alpha \in \{1,\dots,d\}\), and \(\sum_{\alpha=1}^{d} a_{\alpha}^{(i)} = 1\). Hence, \(\mathbf{a}^{(i)} \in \Delta^{d}\), the \(d\)-dimensional probability simplex. For instance, if \(d+1 = 100\), an inorganic material might allow up to 100 distinct species at each site, with \(a_{\alpha}^{(i)}\) capturing the probability of site \(i\) hosting species \(\alpha\).

However, performing a random walk directly on \(\Delta^{d}\) can be tricky. As a component \(a_{\alpha}^{(i)}\)  approaches a boundary, naive step updates risk stepping outside the simplex. To overcome this boundary condition issue, one can first embed \(\Delta^{d}\) into the hypercube \(\mathbb{C}^d=[0,1]^{d}\) and employ modular arithmetic so the random walk wraps around the boundary naturally. The mapping between the two spaces can be done via a {\it bijective} transformation based on uniform spacings \cite{Devroye2013-uniform-spacing}.

Intuitively, we can think of this as mapping a normalized probability (summing to one) onto a stretched counterpart in the hypercube. Then we can perform the walk there with a convenient boundary condition, and then mapping back. Concretely, we convert each \(\mathbf{a}^{(i)}\in\Delta^d\) into a hypercube vector \(\mathbf{A}^{(i)} \in \mathbb{C}^d\), perform geometric random walk in $\mathbb{C}^d$, and finally invert the mapping to identify an updated corresponding probability vector \(\mathbf{a'}^{(i)}\). 

In practice, one might also draw a one-hot site assignment by sampling \(\mathbf{a}^{(i)}\) from a multinomial distribution, effectively picking a vertex of the \(d\)-simplex. Regardless of how these atomic types are sampled, the critical embedding and subsequent inverse transform ensure that the random-walk updates remain in the simplex. Next we describe this bijective transformation.

In the {\it forward} direction, consider a vector \(\mathbf{a} = (a_{1}, a_{2}, \dots, a_{d+1})\) in the \(d\)-simplex \(\Delta^{d}\subset \mathbb{R}^{d+1}\). We construct \(\mathbf{A} = (A_{1}, A_{2}, \dots, A_{d})\) in the \(d\)-dimensional hypercube \(\mathbb{C}^d = [0,1]^{d}\) by setting
\begin{equation}\label{eq:forward_space_trans}
A_{1} = a_{1}, 
\quad
A_{j} = A_{j-1} + a_{j} 
\quad \text{for} \; j = 2, \dots, d.
\end{equation}
We denote this \emph{forward} map of the uniform spacing Eq.~\eqref{eq:forward_space_trans} $\mathcal{H}: \Delta^d \rightarrow \mathbb{C}^d$ as
\begin{equation}
    \mathbf{A} = \mathcal{H}(\mathbf{a}).
\end{equation}

In the {\it inverse} direction, we sort the components of \(\mathbf{A}\) into an ordered list \(\mathbf{z} = (z_{1}, z_{2}, \dots, z_{d})\) from smallest to largest. We then reconstruct \(\mathbf{a}\) by
\begin{equation}\label{eq:hpc_space_trans}
a_{1} = z_{1}, 
\quad
a_{j} = z_{j} - z_{j-1} 
\quad \text{for} \; j = 2, \dots, d,
\quad
a_{d+1} = 1 - z_{d}.
\end{equation}
We define the procedure Eq.~\eqref{eq:hpc_space_trans} as the \emph{inverse} map of the uniform spacing $\mathcal{H}^{-1}: \mathbb{C}^d \rightarrow \Delta^d$ as
\begin{equation}
    \mathbf{a} = \mathcal{H}^{-1}(\mathbf{A}).
\end{equation}

This forward–inverse pair is a bijection, so no information is lost.

Once the atomic-type data live in $\mathbb{C}^d$, we define geodesic updates by a simple modular rule. Suppose we are at a point $\mathbf{x}_1\in \mathbb{C}^d$ and wish to add a tangent step $\mathbf{u}$. We impose a reflecting boundary condition on each component $j$ of the updated vector
\begin{widetext}
\begin{equation}
\label{eq:hyper_expm}
   \left(\mathbf{x}_2\right)_j = \left(\exp_{\mathbf{x}_1}^{\mathbb{C}^d}(\mathbf{u})\right)_j 
   = \newline 
   \begin{cases}
        (\mathbf{x}_1+\mathbf{u})_j \bmod 2, 
        &\text{if }(\mathbf{x}_1+\mathbf{u})_j \bmod 2 < 1,\\
        2 - (\mathbf{x}_1+\mathbf{u})_j \bmod 2, 
        &\text{otherwise}.
    \end{cases}
\end{equation}
\end{widetext}
This reflecting boundary condition enforces the random walk to remain in the $d$-hypercube.

Next, the inverse exponential map under this convention is simply
\begin{equation}
    \mathbf{v}
    \;=\;\exp_{\mathbf{x}_2}^{-1,\mathbb{C}^d}(\mathbf{x}_1) \;=\; \mathbf{x}_1 - \mathbf{x}_2,
\end{equation}
similar to that of the Euclidean manifold. These steps provide a clean way to handle discrete atomic types. 

\begin{figure}[ht]
    \centering
    \includegraphics[width=1\linewidth]{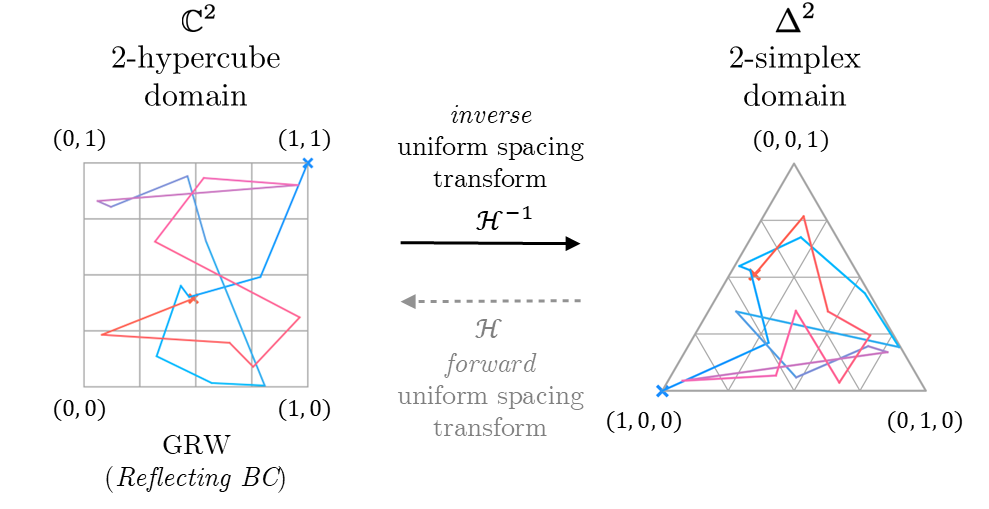}
     \caption{A geodesic random walk (GRW) on a 2-hypercube with a reflecting boundary condition (left), generated using Algorithm~\ref{algo:forward_grw}, is mapped onto a 2-simplex (right) via the hypercube spacing transform ({\it inverse} uniform spacing transform of Eq.~\eqref{eq:hpc_space_trans}). The blue and orange markers indicate the initial and final positions of the walk, respectively.}
    \label{fig:hpc_to_simp}
\end{figure}

Fig.~\ref{fig:hpc_to_simp} illustrates how a geodesic random walk within a 2-hypercube (left) is mapped to a 2-simplex (right) via the hypercube spacing transform. The walk starts near the top-right corner (blue) and finishes around the middle (orange), with each intermediate step residing strictly inside the hypercube. After mapping, every point in the walk lies within the simplex, highlighting that the transformation is bijective and preserves the validity of points. In higher-dimensional settings, the same principle applies; each step on the $d$-hypercube corresponds uniquely to a point in the $d$-simplex.

\section{Equivariant Graph Neural Networks}
\label{appendix:equiformerv2}

Equivariant models are designed so that applying a symmetry transformation to the \emph{input} is equivalent to applying the similar transformation to the \emph{output}. Formally, let $f\colon \mathcal{X} \to \mathcal{Y}$ be a function, and let $g \in G$ be a symmetry operation from some group $G$. We say $f$ is \emph{equivariant} if
\begin{equation}
\label{eq:equivariance}
    f\bigl(g \ast \mathbf{x}\bigr)\;=\;g \ast f(\mathbf{x}), 
    \quad \forall\, g \in G,\;\forall\,\mathbf{x}\in\mathcal{X},
\end{equation}
where ``$\ast$'' denotes the group action on vectors or tensor features. In the case $G=SO(3)$, the group elements can be represented by Wigner $D$-matrices, $\mathbf{D}^l(\mathbf{R})$, indexed by $l$, which capture how tensors of different \emph{ranks} transform under 3D rotations. Concretely, if $\mathbf{x}$ transforms as an $l$-rank feature, then
\begin{equation}
\label{eq:wigner_dmatrix}
    f(\mathbf{x}) 
    \;=\; \mathbf{D}^l\!\bigl(\mathbf{R}^{-1}\bigr)\,\cdot\,
           f\Bigl(\mathbf{D}^l(\mathbf{R})\,\cdot\,\mathbf{x}\Bigr).
\end{equation}
One notable instance of such a function is given by the spherical harmonics $\mathbf{Y}^l(\mathbf{x})$, which exhibit this same symmetry property: applying $\mathbf{D}^l(\mathbf{R}^{-1})$ to the output is equivalent to applying $\mathbf{D}^l(\mathbf{R})$ to the input argument.

\subsection{Equivariant convolutions on graphs}

Building on the idea of preserving $SE(3)$ symmetries (rotations and translations) in neural networks, the Tensor Field Network (TFN)~\cite{Tomas2018tfn} introduced a convolution that propagates features between pairs of nodes while maintaining equivariance. In particular, suppose node $j$ carries a feature $\mathbf{x}_j^{l_i}$ that transforms as an $l_i$-rank tensor, and let $\mathbf{Y}^{l_f}\!\bigl(\hat{\mathbf{r}}_{nj}\bigr)$ encode the geometric dependence on the direction $\hat{\mathbf{r}}_{nj}$ (via spherical harmonics of rank $l_f$). TFN constructs a rank-$l_o$ message by
\begin{equation}
    \label{eq:TFN_message}
    \mathbf{m}_{nj}^{l_o} 
    \;=\; 
    \sum_{l_i,l_f}
    \mathbf{w}_{l_i,l_f,l_o}\,\bigl(\mathbf{x}_j^{l_i} 
          \,\otimes_{l_i,l_f}^{l_o}\, \mathbf{Y}^{l_f}(\hat{\mathbf{r}}_{nj})\bigr),
\end{equation}
where $\otimes_{l_i,l_f}^{l_o}$ is a tensor product mapping $(l_i \times l_f)$ into $l_o$. The learnable weights $\mathbf{w}_{l_i,l_f,l_o}$ specify how each combination of input rank $l_i$ and spherical-harmonic rank $l_f$ contributes to the output rank $l_o$. In practice, these products incorporate Clebsch–Gordan coefficients
\[
   C_{(l_i,m_i),(l_f,m_f)}^{(l_o,m_o)},
\]
which vanish unless $l_f$ lies between $\lvert l_i - l_o\rvert$ and $l_i + l_o$. One can rewrite Eq.~\eqref{eq:TFN_message} more explicitly as
\begin{equation}
    \mathbf{m}_{nj}^{l_o} 
    \;=\; 
    \sum_{l_i,l_f}\mathbf{w}_{l_i,l_f,l_o}\,\bigoplus_{m_o}\Bigl[
       \mathbf{x}_j^{l_i}\,
       \bigl(
         C_{(l_i,m_i),(l_f,m_f)}^{(l_o,m_o)}\,\mathbf{Y}^{l_f}_{m_f}(\hat{\mathbf{r}}_{nj})
       \bigr)
    \Bigr],
\end{equation}
where $\bigoplus_{m_o}$ denotes a concatenation over the index $m_o$. Often, one aligns $\hat{\mathbf{r}}_{nj}$ along a principal axis, simplifying the terms that survive~\cite{Passaro2023escn}.

To handle orientation, a final rotation $(\mathbf{D}^{l_o})^{-1}$ is typically applied. Writing $\tilde{\mathbf{x}}_{j}^{l_i} = \mathbf{D}^{l_o}\mathbf{x}_{j}^{l_i}$, one obtains
\begin{align}
    \label{eq:oriented_msg}
    \mathbf{m}^{l_o}_{nj} 
    \;=\; 
    (\mathbf{D}^{l_o})^{-1}
    \sum_{l_i}\bigoplus_{m_o}\bigl[\mathbf{y}^{l_i,l_o}_{nj}\bigr]_{m_o},
\end{align}
with
\[
\begin{aligned}
    \mathbf{y}^{l_i,l_o}_{m_o,nj} &= \tilde{\mathbf{w}}^{\,l_i,l_o}_{m_o}\,\tilde{\mathbf{x}}_{m_o,j}^{\,l_i} 
                                    \;-\;
                                    \tilde{\mathbf{w}}^{\,l_i,l_o}_{-m_o}\,\tilde{\mathbf{x}}_{-m_o,j}^{\,l_i}
    && \text{for } m_o<0, \\
    \mathbf{y}^{l_i,l_o}_{-m_o,nj} &= \tilde{\mathbf{w}}^{\,l_i,l_o}_{-m_o}\,\tilde{\mathbf{x}}_{m_o,j}^{\,l_i}
                                     \;+\;
                                     \tilde{\mathbf{w}}^{\,l_i,l_o}_{m_o}\,\tilde{\mathbf{x}}_{-m_o,j}^{\,l_i}
    && \text{for } m_o>0, \\
    \mathbf{y}^{l_i,l_o}_{-m_o,nj} &= \tilde{\mathbf{w}}^{\,l_i,l_o}_{m_o}\,\tilde{\mathbf{x}}_{m_o,j}^{l_i}
    && \text{for } m_o=0,
\end{aligned}
\]
where $\tilde{\mathbf{w}}$ are additional learnable parameters.

\subsection{EquiformerV2 and SO2Attn}

Building on TFN-style operations, the Equiformer architecture~\cite{Liao2023equiformer} (and its successor, EquiformerV2~\cite{Liao2024equiformerv2}) integrates \emph{attention} into equivariant message passing. Concretely, scalar features $(l=0)$ undergo a standard attention mechanism, while higher-rank features $(l>0)$ traverse \emph{separable $S^2$ activations} that act on spherical harmonics. 

For scalar features, let $\mathbf{m}_{nj}^{\,l_o=0}$ be the message from node $j$ to node $n$. EquiformerV2 computes an attention score
\[
\begin{aligned}
    \mathbf{z}_{nj} 
    &= \mathbf{w}_a^\top\,\mathrm{LeakyReLU}\bigl(\mathrm{LN}(\mathbf{m}_{nj}^{\,l_o=0})\bigr),\\
    \mathrm{attn}_{nj} 
    &= \frac{\exp(\mathbf{z}_{nj})}{\sum_{k\in \mathcal{N}(n)}\exp(\mathbf{z}_{nk})},
\end{aligned}
\]
where $\mathrm{LN}(\cdot)$ denotes layer normalization, and $\mathrm{LeakyReLU}$ is a standard nonlinear activation. These $\mathrm{attn}_{nj}$ scalars dictate how strongly node $j$ influences node $n$. Meanwhile, any $l_o>0$ features $\mathbf{m}_{nj}^{\,l_o}$ are mapped to $S^2$ point samples, processed by a nonlinearity (e.g.\ SiLU), and then reconverted to irreps, ensuring full rotational equivariance. EquiformerV2 multiplies the scalar and higher-rank pathways to combine geometric and scalar information, before converting the result back to the original frame via $(\mathbf{D}^{l_o})^{-1}$. The final node update for a rank-$l_o$ feature becomes
\begin{equation}
\label{eq:output_head}
    \mathbf{h}_n^{l_o}
    \;=\;
    \mathrm{Linear} \Bigl(
      \sum_{j \in \mathcal{N}(n)} \mathbf{m}_{nj}^{\,l_o}
    \Bigr),
\end{equation}
where $\mathrm{Linear}(\cdot)$ is a fully connected layer. The combination of TFN-style convolutions and attention-based $S^2$ activations is termed \emph{SO2Attn} in EquiformerV2, enabling fully equivariant message passing for 3D geometric data.

By adopting these equivariant operations, EquiformerV2 preserves rotational and translational symmetries in materials or molecular graphs, yielding embeddings or predictions (e.g.\ forces, energies, or diffusion-model scores) that remain consistent under global $SE(3)$ transformations, a crucial inductive bias for crystalline materials modeling.

\noindent{\bf Output heads} 
After obtaining the hidden features from Eq.~\eqref{eq:output_head}, CrystalGRW, by using EquiformerV2, generates three outputs corresponding to each manifold slice:
\[
\begin{aligned}
    \mathbf{s}_{\theta}^{\mathbb{T}^3} &= \mathrm{SO2Attn}\bigl(\mathbf{h}_n^{l_o=1}\bigr),\\
    \mathbf{s}_{\theta}^{\mathbb{C}^d} &= \mathrm{SO2Attn}\bigl(\mathbf{h}_n^{l_o=0}\bigr),\\
    \mathbf{s}_{\theta}^{\mathbb{R}^{3\times 3}} &= \mathrm{global\_pool}\bigl(\mathrm{FFN}\bigl(\mathbf{h}_n^{l_o=0}\bigr)\bigr),
\end{aligned}
\]
where $\mathrm{FFN}$ denotes a feed-forward network and global\_pool is a function that pools node-level features into graph-level features. This setup allows the network to produce manifold-specific scores or feature vectors for fractional coordinates, atomic types, and lattice parameters, respectively. 
Specifically, these three heads  then form the overall score function, 
\begin{equation}
\label{eq:score_heads}
    \mathbf{s}_\theta(\mathbf{x}_t, t)
  \;=\;
  \left(
    \mathbf{s}_{\theta}^{\mathbb{T}^3},\;
    \mathbf{s}_{\theta}^{\mathbb{C}^d},\;
    \mathbf{s}_{\theta}^{\mathbb{R}^{3\times 3}}
  \right),
\end{equation}
so that each manifold slice receives its own specialized output.

\section{Score Function for Atomic Types}
\label{sec:appendix_atomic_types}
\subsection{Atomic-type estimator}
For atomic-type variables, recall the goal to learn a score function $\mathbf{s}^{\mathbb{C}^d}_{\theta}(\mathbf{a}^{(i)}_t,t)$ that, at each time $t$, points from the noised $\mathbf{A}^{(i)}_t$ toward the original atomic-type configuration $\mathbf{A}^{(i)}_0$. In principle, we could approximate this score by learning the manifold inverse exponential map directly from 
\[
   \mathbf{s}^{\mathbb{C}^d, (i)}_{\theta}(\mathbf{a}^{(i)}_t,t)
   \;\approx\; 
   \frac{1}{t}\,\exp_{\mathbf{A}^{(i)}_t}^{-1, {\mathbb{C}^d}}\!\bigl(\mathbf{A}^{(i)}_0\bigr).
\]
However, an alternative that perform better at sampling atomic types in our experiments is to introduce a network $\mathbf{G}^{\Delta^d}_{\theta}(\mathbf{a}^{(i)}_t,t)$, which directly predicts $\mathbf{\hat{a}}^{(i)}_0$ from $\mathbf{a}^{(i)}_t$ by
\begin{equation}
\begin{aligned}
    \mathbf{\hat{a}}^{(i)}_0 &= \mathbf{G}^{\Delta^d}_{\theta}\bigl(\mathbf{a}^{(i)}_t,t\bigr), \\
    \mathbf{\hat{A}}^{(i)}_0 &= \mathcal{H}\bigl(\mathbf{\hat{a}}^{(i)}_0\bigr),
\end{aligned}
\end{equation}
where $\mathbf{\hat{a}}^{(i)}_0$ and $\mathbf{\hat{A}}^{(i)}_0$ are estimates of original data in $\Delta^d$ and $\mathbb{C}^d$, respectively.
The score can then be recovered by composing the network with the inverse exponential map by
\begin{equation}
\label{eq:atom_type_loss}
    \mathbf{s}^{\mathbb{C}^d, (i)}_{\theta}\bigl(\mathbf{a}^{(i)}_t,t\bigr)
    \;=\;
    \frac{1}{t}\,
\exp_{\mathbf{A}^{(i)}_t}^{-1, {\mathbb{C}^d}} \!\Bigl(\mathcal{H}\bigl(\mathbf{G}^{\Delta^d}_{\theta}(\mathbf{a}^{(i)}_t,t)\bigr)\Bigr).
\end{equation}

In this approach, the network $\mathbf{G}^{\Delta^d}_{\theta}$ is trained via a mean-squared error criterion that compares its prediction $\mathbf{G}^{\Delta^d}_{\theta}(\mathbf{a}^{(i)}_t,t)$ to the ground truth $\mathbf{a}^{(i)}_0$
\begin{equation}
\label{eq:loss_atom}
  \mathcal{L}^{\Delta^d}_\theta
    \;=\;
    \frac{\lambda_{\Delta^d}}{N}
    \sum_{i=1}^{N}
       \bigl\|\,
         \mathbf{G}^{\Delta^d}_{\theta}\bigl(\mathbf{a}^{(i)}_t,t\bigr)
         \;-\;
         \mathbf{a}^{(i)}_0
       \bigr\|^{2}.
\end{equation}
After training, we can switch back to the score-based perspective by applying Eq.~\eqref{eq:atom_type_loss} whenever we need the gradient of the log-probability with respect to $\mathbf{A}^{(i)}_t$. 

In the actual implementation, CrystalGRW utilizes Algorithms~\ref{algo:train_real} and \ref{algo:reverse_grw_real} for training and sampling, respectively. The model's output head for generating atomic types is
\[
\begin{aligned}
\mathbf{G}_{\theta}^{\Delta^d} &= \mathrm{softmax}\left(\mathrm{SO2Attn}\bigl(\mathbf{h}_n^{l_o=0}\bigr)\right),
\end{aligned}
\]
thereby giving the overall score function
\begin{equation}
    \mathbf{s}_\theta(\mathbf{x}_t, t) = \left(\mathbf{s}_{\theta}^{\mathbb{T}^3}, \mathbf{G}_{\theta}^{\Delta^d}, \mathbf{s}_{\theta}^{\mathbb{R}^{3\times 3}}\right).
\end{equation}

\subsection{Comparisons of Model Variants for Perovskite Compositions}

\begin{figure}[t]
    \centering
    \includegraphics[width=0.85\linewidth]{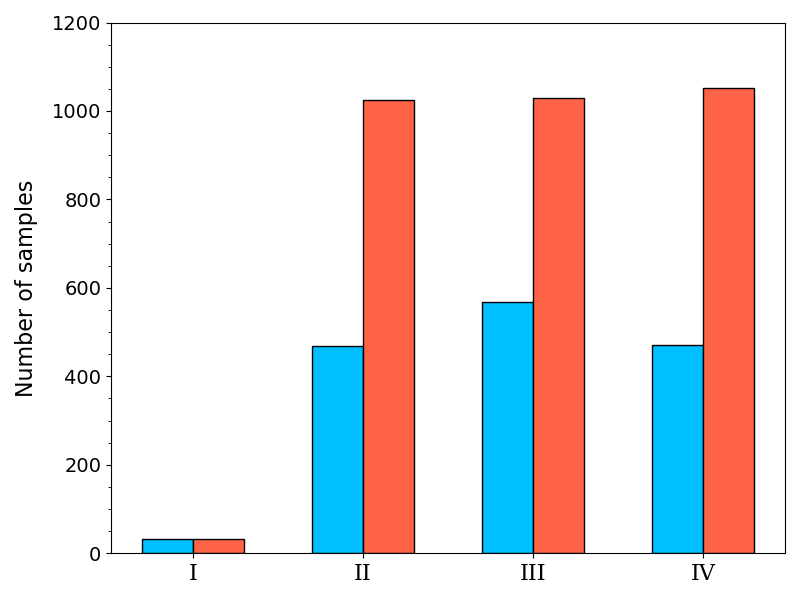}
    \caption{Number of generated structures (from 1280 samples) identified as having perovskite compositions across four model variants: Model~I predicts $\mathbf{s}_{\theta}^{\mathbb{C}^d}$, Models~II and III predict $\mathbf{G}^{\Delta^d}_{\theta}$ where output heads are FFN and SO2Attn, respectively, and Model~IV is similar to Model~II but trained with the cross-entropy loss. Red bars indicate the total count of perovskite structures produced, while blue bars represent how many are unique within each method.}
    \label{fig:atom_type_perform}
\end{figure}

Fig.~\ref{fig:atom_type_perform} compares four approaches for predicting perovskite compositions where the models are trained with the Perov-5 dataset \cite{Xie2022cdvae}. Model~I employs a scheme where the network predicts all three scores as in Eq.~\eqref{eq:score_heads}. Models~II and~III rely on a model's architecture (Eq.~\eqref{eq:atom_type_loss}), differing in their neural network heads (feed-forward vs.\ SO2Attn) while their model's schemes are to predict $\mathbf{s}_\theta(\mathbf{x}_t, t) = \left(\mathbf{s}_{\theta}^{\mathbb{T}^3}, \mathbf{G}^{\Delta^d}_{\theta}, \mathbf{s}_{\theta}^{\mathbb{R}^{3\times 3}}\right)$. Model~IV is similar to Model~II but replaces the mean-squared error with a cross-entropy objective. Although Model~I struggles to generate valid perovskite structures, Models~II, III, and IV perform substantially better in capturing the correct stoichiometry. Among these, Model~III yields the highest fraction of unique ones, so our choice of architecture for atomic-type head is SO2Attn.

\begin{algorithm}[t]
\noindent\hrulefill\par\nobreak\vskip-10pt
\caption{\centering Modified \emph{Training}  Multi-Manifold Crystal Properties}
\vskip-7pt\hrulefill
\label{algo:train_real}
\begin{algorithmic}[1]

\REQUIRE \Statex 
\begin{itemize}
    \item $\mathbf{x}_0 =\bigl(\mathbf{\tilde{r}}_0,\;\mathbf{\tilde{a}}_0,\;\mathbf{L}_0\bigr)$, each on manifolds $\left(\mathbb{T}^3\right)^N, \left(\Delta^d\right)^N, \mathbb{R}^{3\times3}$ respectively

    \item Total steps $K$
\end{itemize}
\vspace{5pt}  

\State $\mathbf{\tilde{A}}_0 = \mathcal{H}(\mathbf{\tilde{a}}_0)$
\State $\mathbf{X}_0 =\bigl(\mathbf{\tilde{r}}_0,\;\mathbf{\tilde{A}}_0,\;\mathbf{L}_0\bigr)$

\REPEAT
    \State \(t \sim  \mathrm{Uniform}(\eta, T)\) 
    \State \(\bigl(\mathbf{\tilde r}_t,\;\mathbf{\tilde A}_t,\;\mathbf{L}_t\bigr) \gets \mathrm{GRW}\left(\mathbf{X}_0, K, t\right)\) \Rulecomment{Algorithm \ref{algo:forward_grw}}
    \State Compute inverse exponential maps
    \[
    \begin{aligned}
      \mathbf{v}^{\mathbb{T}^3, (i)}_t \;&=\;
\exp_{\mathbf{r}_t^{(i)}}^{-1,\mathbb{T}^3}\bigl(\mathbf{r}_0^{(i)}\bigr),\\
\mathbf{v}^{\mathbb{C}^d, (i)}_t \;&=\;
\exp_{\mathbf{A}_t^{(i)}}^{-1,\mathbb{C}^d}\bigl(\mathbf{A}_0^{(i)}\bigr),\\
\mathbf{v}^{\mathbb{R}^{3\times 3}}_t \;&=\;
\exp_{\mathbf{L}_t}^{-1,\mathbb{R}^{3\times 3}}\bigl(\mathbf{L}_0\bigr)
    \end{aligned}
  \]
\State $\mathbf{\tilde{a}}_t = \mathcal{H}^{-1}(\mathbf{\tilde{A}}_t)$
\State $\mathbf{x}_t =\bigl(\mathbf{\tilde{r}}_t,\;\mathbf{\tilde{a}}_t,\;\mathbf{L}_t\bigr)$
\State $\bigl(\mathbf{s}^{\mathbb{T}^3}_{\theta}, \mathbf{G}^{\Delta^d}_{\theta}, \mathbf{s}^{\mathbb{R}^{3\times 3}}_{\theta}\bigr) \gets \mathbf{s}_{\theta}(\mathbf{x}_t,t)$

\State Compute the loss  \Rulecomment{Varadhan's asymptotic}
    \[
    \begin{aligned}
      \mathcal{L}_\theta
      &\;=\;
      \frac{\lambda_{(\mathbb{T}^3)^N}}{N} \sum_{i=1}^N\Bigl\|
        \mathbf{s}_\theta^{\mathbb{T}^3, (i)}
        \;-\;
        \mathbf{v}^{\mathbb{T}^3, (i)}_t/t
      \Bigr\|^{2}
      \\
      &\;+\;
      \frac{\lambda_{\Delta^d}}{N} \sum_{i=1}^N\Bigl\|
        \mathbf{G}^{\Delta^d, (i)}_\theta
        \;-\;
        \mathbf{a}^{(i)}_0
      \Bigr\|^{2}\\
      &\;+\;
      \lambda_{\mathbb{R}^{3\times 3}} \Bigl\|
        \mathbf{s}_\theta^{\mathbb{R}^{3\times 3}}
        \;-\;
        \mathbf{v}^{\mathbb{R}^{3\times 3}}_t/t
      \Bigr\|^{2}
    \end{aligned}
  \]

    \State  Take gradient descent step on \( \mathbf{\nabla}_{\theta} \mathcal{L}_{\theta} \)
\UNTIL{converged}
\end{algorithmic}
\end{algorithm}

\begin{algorithm}[h]
\noindent\hrulefill\par\nobreak\vskip-10pt
\caption{\centering Modified Reverse Geodesic Random Walk (\emph{Sampling}) for Multi-Manifold States}
\vskip-7pt\hrulefill
\label{algo:reverse_grw_real}
\begin{algorithmic}[1]

\REQUIRE \(N, K, T, \xi, \mathbf{s}_{\theta}\)
\vspace{5pt}

\State $\mathbf{\tilde{r}}_T = (\mathbf{r}_T^{(1)}, \dots, \mathbf{r}_T^{(N)}) \sim \mathrm{Uniform}\bigl([0,1)\bigr)^{N \times 3}$ 

\State $\mathbf{\tilde{A}}_T = (\mathbf{A}_T^{(1)}, \dots, \mathbf{A}_T^{(N)})  \sim \text{Uniform}\left([0,1]^d\right)^N$  

\State $\mathbf{L}_T \sim \mathcal{N}\!\Bigl(\mathbf{0},\,\tfrac{h_f + h_0}{2}\,\mathbf{I}\Bigr)$ where $\mathbf{0}, \mathbf{I}\in \mathbb{R}^{3\times 3}$

\State $\mathbf{\tilde a}_{K} = \mathcal{H}^{-1}(\mathbf{\tilde A}_{K})$

  \State $\mathbf{x}_{K} = (\mathbf{\tilde r}_{K},\;\mathbf{\tilde a}_{K},\;\mathbf{L}_{K})$

\FOR{$k = K \text{ to }  1$}
  \State $\gamma = (T/K)k^{\xi}$
  \State $t = \gamma k$
  
  \State $\bigl(\mathbf{s}^{\mathbb{T}^3}_{\theta}, \mathbf{G}^{\Delta^d}_{\theta}, \mathbf{s}^{\mathbb{R}^{3\times 3}}_{\theta}\bigr) \gets \mathbf{s}_{\theta}(\mathbf{x}_k, t)$
  \State
  $\mathbf{s}^{\mathbb{C}^d, (i)}_{\theta}
    \;=\;
    \frac{1}{t}\exp_{\mathbf{A}^{(i)}_t}^{-1, {\mathbb{C}^d}} \!\left(\mathcal{H}\bigl(\mathbf{G}^{\Delta^d, (i)}_{\theta}\bigr)\right)$
  \State Get drift coefficients
        \[
        \bigl(\underbrace{\mathbf{0}}_{\left(\mathbb{T}^3\right)^N},
            \underbrace{\mathbf{0}}_{\left(\mathbb{C}^d\right)^N},
            \underbrace{\mathbf{b}^{\mathbb{R}^{3\times 3}}}_{\approx \mathbf{0}\in\mathbb{R}^{3\times3}}\bigr)
        \gets \mathbf{b}\bigl(\mathbf{X}_{k}, t\bigr)
       \]

  \State Apply to all manifolds
  \[
  \begin{aligned}
  \mathbf{W}^{\mathcal{M}}_{k-1} = \gamma\Bigl[-\mathbf{b}^{\mathcal{M}} +\mathbf{s}^{\mathcal{M}}_{\theta}
                 \Bigr] + \sqrt{\gamma}\sigma^{\mathcal{M}}(t)\bm{\epsilon}_k, \\ \bm{\epsilon}_k \sim \mathcal{N}\bigl(\mathbf{0},\mathbf{I}\bigr)
  \end{aligned}
  \]
  
  \State $(\mathbf{\tilde r}_k,\,\mathbf{\tilde A}_k,\,\mathbf{L}_k) \gets \mathbf{X}_{k}$
  \State Apply manifold-specific exponential maps
  \[
  \begin{aligned}
    \mathbf{r}_{k-1}^{(i)} &= \exp_{\mathbf{r}_{k}^{(i)}}^{\mathbb{T}^3}\!\bigl(\mathbf{W}_{k-1}^{\mathbb{T}^3, (i)}\bigr), \\
    \mathbf{A}_{k-1}^{(i)} &= \exp_{\mathbf{A}_{k}^{(i)}}^{\mathbb{C}^d}\!\bigl(\mathbf{W}_{k-1}^{\mathbb{C}^d, (i)}\bigr), \\
    \mathbf{L}_{k-1} &= \exp_{\mathbf{L}_{k}}^{\mathbb{R}^{3\times 3}}\!\bigl(\mathbf{W}_{k-1}^{\mathbb{R}^{3\times 3}}\bigr)
  \end{aligned}
  \]
  \State $\mathbf{\tilde a}_{k-1} = \mathcal{H}^{-1}(\mathbf{\tilde A}_{k-1})$

  \State $\mathbf{x}_{k-1} = (\mathbf{\tilde r}_{k-1},\;\mathbf{\tilde a}_{k-1},\;\mathbf{L}_{k-1})$
\ENDFOR
\State \textbf{return} $\mathbf{x}_0 = (\mathbf{\tilde r}_0,\;\mathbf{\tilde a}_0,\;\mathbf{L}_0)$

\end{algorithmic}
\end{algorithm}

\section{Point group control}
\label{sec:pg_control}

\begin{figure}[ht]
    \centering
    \includegraphics[width=\linewidth]{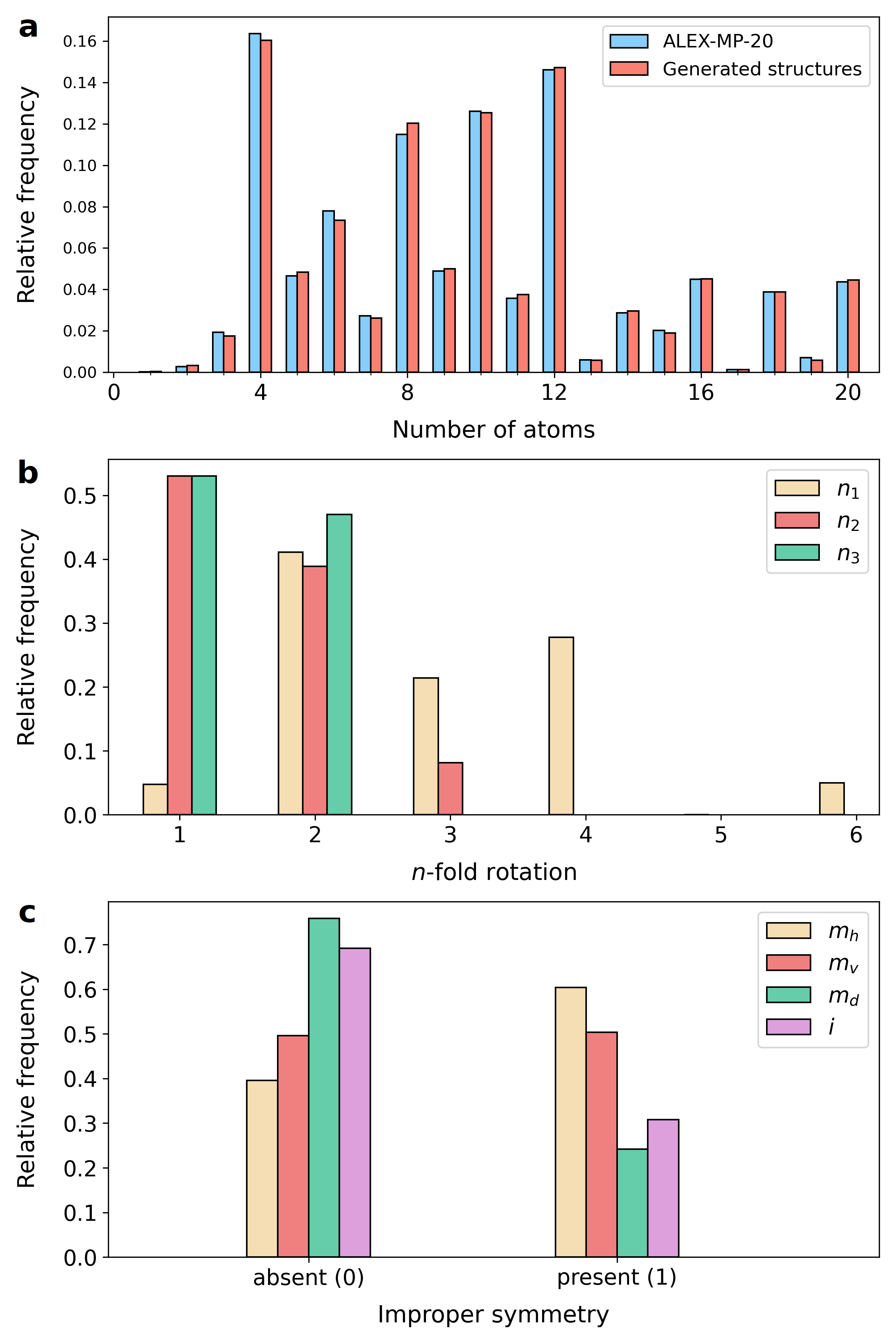}
    \caption{(a) Relative frequency of the number of atoms in a unit cell for the ALEX-MP-20 dataset and 10$^4$ generated structures. (b) Relative frequency of the three principal rotation-axis classes ($n_1,n_2,n_3$). (c) Fraction of structures for which each of the three mirror-plane types ($m_h,m_v,m_d$) and inversion symmetry ($i$) is present or absent.}
    \label{fig:alexmp20_stats}
\end{figure}

Point groups in the ALEX-MP-20 dataset are imbalanced (Fig.~\ref{fig:pg_gen_alexmp20}). For example, trigonal and hexagonal lattices (e.g.\ $\bar{3}$ or $6mm$) are far less frequent than cubic or orthorhombic lattices. To address this, we aggregate the 32 point groups into seven classes covering three principal rotation-axis classes ($n_1,n_2,n_3$), three mirror-plane classes ($m_h,m_v,m_d$), and inversion symmetry ($i$). Figures~\ref{fig:alexmp20_stats}(b) and (c) show how these classes appear in MP-20: panel (b) focuses on the distribution of the three principal rotation-axis classes, while panel (c) indicates whether each mirror-plane class and inversion are present or absent. Together, these aggregated classes are more evenly represented than the original 32 point groups.

We set $n_1$ with 6 possible classes, ranging from 1-fold to 6-fold (although 5-fold does not exist for periodic crystals). For $n_2$, the possible folds are 1, 2, and 3, assigning it 3 classes, while $n_3$ has folds up to 2-fold, assigning it 2 classes. Mirror planes and inversion symmetry have possible classes of 0 or 1, indicating their absence or presence, respectively. For $m\bar{3}m$, it has 4-, 3-, 2-rotation axes, all the mirror planes, and inversion symmetry, so its labels are 3, 2, 1, 1, 1, 1, 1 for $n_1$, $n_2$, $n_3$, $m_h$, $m_v$, $m_d$, $i$, respectively. For $432$, it has 4-, 3-, 2-rotation axes without improper symmetry, so its labels are 3, 2, 1, 0, 0, 0, 0. For $6/mmm$, it has 6-, 2-, 2-rotation axes and $m_h$, $m_v$ planes, so its labels are 5, 1, 1, 1, 1, 0.  For $\bar{1}$, its rotation axes are 1-fold, and it has inversion symmetry, assigning it the labels 0, 0, 0, 0, 0, 0, 1. Labels of each point group are displayed in Table~\ref{tab:pg_labels}.

Point groups of output structures are controlled using seven labels. Each label is one-hot encoded and passed through two fully connected layers; the resulting embeddings are treated as node attributes in the graph neural network. For classifier-free guidance, the model is trained both with point group conditions and with a null condition. The null condition is applied to randomly selected samples (with a probability of 0.1) by zeroing out the embedded features.

During inference, the point group is controlled by selecting seven labels corresponding to the target symmetry, as previously described. The score conditioned on the point group is computed using Eq.~\eqref{eq:conditional_score}, with $w = 0.5$.

\begin{table}[h]
\centering
\caption{Labels of 32 point groups for conditional control.}
\begin{ruledtabular}
\begin{tabular}{cccccccc}
    Point group & $n_1$ & $n_2$ & $n_3$ & $m_h$ & $m_v$ & $m_d$ & $i$ \\
    \hline
    $1$ & 0 & 0 & 0 & 0 & 0 & 0 & 0  \\
    $\bar{1}$ & 0 & 0 & 0 & 0 & 0 & 0 & 1  \\
    $2$ & 1 & 0 & 0 & 0 & 0 & 0 & 0  \\
    $m$ & 0 & 0 & 0 & 1 & 0 & 0 & 0 \\
    $3$ & 2 & 0 & 0 & 0 & 0 & 0 & 0 \\
    $2/m$ & 1 & 0 & 0 & 1 & 0 & 0 & 0 \\
    $222$ & 1 & 1 & 1 & 0 & 0 & 0 & 0 \\
    $mm2$ & 1 & 0 & 0 & 0 & 1 & 0 & 0 \\
    $4$ & 3 & 0 & 0 & 0 & 0 & 0 & 0 \\
    $\bar{4}$ & 3 & 0 & 0 & 0 & 0 & 0 & 1  \\
    $\bar{3}$ & 2 & 0 & 0 & 0 & 0 & 0 & 1 \\
    $32$ & 2 & 1 & 1 & 0 & 0 & 0 & 0 \\
    $3m$ & 2 & 0 & 0 & 0 & 1 & 0 & 0 \\
    $6$ & 5 & 0 & 0 & 0 & 0 & 0 & 0 \\
    $\bar{6}$ & 5 & 0 & 0 & 0 & 0 & 0 & 1 \\
    $mmm$ & 1 & 1 & 1 & 1 & 1 & 0 & 0 \\
    $4/m$ & 3 & 0 & 0 & 1 & 0 & 0 & 0 \\
    $422$ & 3 & 1 & 1 & 0 & 0 & 0 & 0 \\
    $4mm$ & 3 & 0 & 0 & 0 & 1 & 0 & 0 \\
    $\bar{4}2m$ & 3 & 1 & 1 & 0 & 0 & 1 & 1 \\
    $\bar{3}m$ & 2 & 0 & 0 & 0 & 0 & 1 & 1 \\
    $6/m$ & 5 & 0 & 0 & 1 & 0 & 0 & 0 \\
    $622$ & 5 & 1 & 1 & 0 & 0 & 0 & 0 \\
    $6mm$ & 5 & 0 & 0 & 0 & 1 & 0 & 0 \\
    $\bar{6}m2$ & 5 & 1 & 1 & 0 & 1 & 0 & 1 \\
    $23$ & 2 & 2 & 1 & 0 & 0 & 0 & 0 \\
    $4/mmm$ & 3 & 1 & 1 & 1 & 1 & 0 & 0 \\
    $6/mmm$ & 5 & 1 & 1 & 1 & 1 & 0 & 0 \\
    $m\bar{3}$ & 2 & 1 & 1 & 1 & 1 & 0 & 1 \\
    $432$ & 2 & 2 & 1 & 0 & 1 & 1 & 1 \\
    $\bar{4}3m$ & 2 & 2 & 1 & 0 & 1 & 1 & 1 \\
    $m\bar{3}m$ & 3 & 2 & 1 & 1 & 1 & 1 & 1 \\
\end{tabular}
\end{ruledtabular}
\label{tab:pg_labels}
\end{table}

\section{Step-Length Scheduler on a General Manifold}
\label{sec:appendix_sde}

\subsection{SDE Formulation and Score-Based Generative Modeling}

Consider a stochastic process \(\mathbf{x}_t \in \mathcal{M}\), where \(\mathcal{M}\) is an arbitrary Riemannian manifold. In local coordinates, an SDE on \(\mathcal{M}\) can be written as
\begin{equation}
    \label{eq:sde_general}
    d\mathbf{x}_t
    \;=\;
    \mathbf{b}(\mathbf{x}_t, t)\,dt 
    \;+\;
    \sigma(\mathbf{x}_t, t)\,d\mathbf{B}_t^{\mathcal{M}},
\end{equation}
where \(\mathbf{B}_t^{\mathcal{M}}\) denotes Brownian motion intrinsic to \(\mathcal{M}\), \(\mathbf{b}\) is a \emph{drift} vector field on the tangent bundle $\mathcal{T}_{\mathbf{x}_t}\mathcal{M}$, and \(\sigma\) is a scalar (or tensor) diffusion coefficient. In many \emph{score-based generative models}, one wishes to design \(\mathbf{b}\) and \(\sigma\) so that, by time \(T\), samples \(\mathbf{x}_T\) approximate some simple “noise” distribution (e.g.\ uniform on a compact manifold or wide Gaussian on a non-compact manifold).

\paragraph{Forward and Reverse SDE.}
When data \(\mathbf{x}_0\sim p_0(\mathbf{x})\) is corrupted forward, \(\mathbf{x}_t\) evolves from \(t=0\) to \(t=T\).  Reversing this process amounts to sampling from \(\mathbf{x}_T\sim \rho_T\) and integrating backward in time where $\rho_T$ is a prior distribution.  Formally, if we pick local coordinates and treat
\[
    \mathbf{b}(\mathbf{x},t) 
    \;=\; f(t)\,\mathbf{x}, 
    \quad
    \sigma(\mathbf{x},t) 
    \;=\; g(t),
\]
one obtains an SDE akin to
\begin{equation}
    \label{eq:forward_sde_simple}
    d\mathbf{x}
    \;=\;
    f(t)\,\mathbf{x}\,dt 
    \;+\;
    g(t)\,d\mathbf{B}_t^\mathcal{M}.
\end{equation}
Instead of Anderson time reversal SDE \cite{Anderson1982}, RSGM uses F\"ollmer time reversal SDE \cite{Follmer1985, Cattiaux2023} as
\begin{equation}
\label{eq:reverse_sde_general}
    d\mathbf{x}
    \;=\;
    \Bigl[
        -\,f(t)\,\mathbf{x} 
        \;+\;
        \,\nabla_{\mathbf{x}}\!\log p_t(\mathbf{x})
    \Bigr]\,dt
    \;+\;
    g(t)\,d\widetilde{\mathbf{B}}_t^\mathcal{M},
\end{equation}
where \(\nabla_{\mathbf{x}}\!\log p_t(\mathbf{x})\) is the \emph{score} of the distribution at time \(t\) (defined intrinsically on the manifold), and \(\widetilde{\mathbf{B}}_t^\mathcal{M}\) is the backward Brownian motion.  Score-based generative models learn a network \(s^{\mathcal{M}}_\theta(\mathbf{x},t)\approx \nabla_{\mathbf{x}}\!\log p_t(\mathbf{x})\).  

\paragraph{Score-Matching and Varadhan’s Asymptotic.}

We begin with the forward diffusion on a manifold \(\mathcal{M}\). Let \(p_t(\mathbf{x})\) be the density of \(\mathbf{x}_t\) at time \(t\). Recall we seek to learn a score function
\[
  s_\theta^{\mathcal{M}}(\mathbf{x},t)\;\approx\;\nabla_{\mathbf{x}}\log p_t(\mathbf{x}),
\]
by minimizing the expected mean-squared difference
\begin{equation}
  \label{eq:score_matching_loss_app}
  \mathcal{L}_t
  \;=\;
  \mathbb{E}_{\mathbf{x}\sim p_t}\!\Bigl[
    \bigl\|\,
      s_\theta^{\mathcal{M}}(\mathbf{x},t)
      \;-\;
      \nabla_{\mathbf{x}}\log p_t(\mathbf{x})
    \bigr\|^{2}
  \Bigr].
\end{equation}

To connect \(\nabla_{\mathbf{x}}\log p_t(\mathbf{x})\) with the original data \(\mathbf{x}_0\), wefirst rewrite
\[
  p_t(\mathbf{x})
  \;=\;
  \int_{\mathcal{M}} p_t\bigl(\mathbf{x}\mid \mathbf{x}_0\bigr)\,p_0(\mathbf{x}_0)\,d\mathbf{x}_0,
\]
where \(p_t(\mathbf{x}\mid \mathbf{x}_0)\) is the \emph{forward transition} kernel describing how \(\mathbf{x}_0\) evolves to \(\mathbf{x}_t\). We can rewrite
\[
  \nabla_{\mathbf{x}}\log p_t(\mathbf{x})
  \;=\;
  \nabla_{\mathbf{x}}
  \log
  \Bigl(
    \int_{\mathcal{M}} p_t(\mathbf{x}\mid \mathbf{x}_0)\,p_0(\mathbf{x}_0)\,d\mathbf{x}_0
  \Bigr).
\]
By applying a small-time (or local) expansion to \(p_t(\mathbf{x}\mid \mathbf{x}_0)\) for \(t\) near zero, or by using a Girsanov's theorem to relate forward and reverse SDEs (cf.\ \cite{Bortoli2022rsgm}), one obtains the approximation
\[
  \nabla_{\mathbf{x}}\log p_t(\mathbf{x})
  \;\approx\;
  \frac{1}{t}\,\exp_{\mathbf{x}}^{-1,\mathcal{M}}\bigl(\mathbf{x}_0\bigr)
\]
for \(\mathbf{x}\) close to \(\mathbf{x}_0\). Intuitively, for very small \(t\), the diffusion around \(\mathbf{x}_0\) is almost Gaussian-like in a local coordinate patch, and \(\exp_{\mathbf{x}}^{-1,\mathcal{M}}(\mathbf{x}_0)\) can be thought of as the vector in \(\mathcal{T}_{\mathbf{x}}\mathcal{M}\) pointing from \(\mathbf{x}\) to \(\mathbf{x}_0\). This leads to \emph{Varadhan’s asymptotic} \cite{Varadhan1967}, which states that
\[
  p_t\bigl(\mathbf{x}\mid \mathbf{x}_0\bigr)
  \;\sim\;
  \exp\Bigl[-\,\tfrac{d_{\mathcal{M}}^2(\mathbf{x},\mathbf{x}_0)}{2\,t}\Bigr]
  \quad (\text{as } t\to 0),
\]
where \(d_{\mathcal{M}}(\mathbf{x},\mathbf{x}_0)\) is the geodesic distance. Differentiating with respect to  \(\mathbf{x}\) gives
\[
  \nabla_{\mathbf{x}}\log p_t\bigl(\mathbf{x}\mid \mathbf{x}_0\bigr)
  \;\approx\;
  \frac{\exp_{\mathbf{x}}^{-1,\mathcal{M}}(\mathbf{x}_0)}{t}.
\]
Hence, in the score-matching objective Eq.~\eqref{eq:score_matching_loss_app}, one can replace the true \(\nabla_{\mathbf{x}}\!\log p_t(\mathbf{x})\) by \(\tfrac{1}{t}\,\exp_{\mathbf{x}}^{-1,\mathcal{M}}(\mathbf{x}_0)\) in small-$t$ regimes or within each noising step. 

When implementing the forward diffusion in discrete time, one typically draws a random $t$ from \([\eta,T]\), applies a geodesic random walk from \(\mathbf{x}_0\) to \(\mathbf{x}_t\), and then enforces
\[
  s_\theta^{\mathcal{M}}\bigl(\mathbf{x}_t,t\bigr)
  \;\approx\;
  \frac{1}{t}\,\exp_{\mathbf{x}_t}^{-1,\mathcal{M}}\!\bigl(\mathbf{x}_0\bigr).
\]
Minimizing
\[
  \mathcal{L}_t
  \;=\;
  \mathbb{E}_{\mathbf{x}_0,\,\mathbf{x}_t}\Bigl[
    \bigl\|\,
      s_\theta^{\mathcal{M}}(\mathbf{x}_t,t)
      \;-\;
      \tfrac{1}{t}\,\exp_{\mathbf{x}_t}^{-1,\mathcal{M}}(\mathbf{x}_0)
    \bigr\|^2
  \Bigr]
\]
ensures $s_\theta^{\mathcal{M}}$ learns to approximate the manifold score. Empirical results \cite{Bortoli2022rsgm} show that this approach still accurately reconstructs the data distribution, even though the local expansion is formally exact only as $t \to 0$. Also, note that Girsanov’s theorem \cite{Girsanov1960, Karatzas1998} formally justifies rewriting the SDE in reverse time, while the local expansions (Varadhan’s asymptotic) give the explicit \(\frac{1}{t}\exp_{\mathbf{x}_t}^{-1,\mathcal{M}}(\mathbf{x}_0)\) form.  For additional details and proofs in a Riemannian context, see \cite{Bortoli2022rsgm} or \cite{Arnaudon2014browniangeo} and the references therein for manifold-based diffusions.

\subsection{Fokker--Planck Equation and Closed-Form Solutions}

In local coordinates, the forward density \(p(\mathbf{x},t\mid \mathbf{x}_0)\) on a Riemannian manifold \(\mathcal{M}\) satisfies a Fokker--Planck (Kolmogorov forward) equation of the form
\begin{align*}
\label{eq:fpe_general}
    \frac{\partial}{\partial t}\,
    p(\mathbf{x},t\mid \mathbf{x}_0)
    = &-\,\mathrm{div}\Bigl[
       p(\mathbf{x},t\mid \mathbf{x}_0)\,\mathbf{b}(\mathbf{x},t)
    \Bigr]\\
    &+\;\tfrac12\,\mathrm{trace}\Bigl[
       \nabla^2 \Bigl(
         p(\mathbf{x},t\mid \mathbf{x}_0)\,\Sigma(\mathbf{x},t)
       \Bigr)
    \Bigr],
\end{align*}
where \(\mathbf{b}(\mathbf{x},t)\) is the drift in local coordinates, and \(\Sigma(\mathbf{x},t)\) is the \emph{diffusion tensor}. If the noise amplitude is \(\sigma(\mathbf{x},t)\) and the metric tensor is \(g_{ij}(\mathbf{x})\), then one often writes
\[
    \Sigma(\mathbf{x},t)
    \;=\;
    \sigma(\mathbf{x},t)^2\,g^{ij}(\mathbf{x}),
\]
so that \(\Sigma\) is a matrix specifying how Brownian increments are curved by \(g_{ij}\). In fact, the corresponding generator of this SDE is closely related to the Laplace--Beltrami operator \(\Delta\), which generalizes the flat-space \(\nabla^2\). On a \emph{compact} manifold, one frequently sets \(\mathbf{b}(\mathbf{x},t)=\mathbf{0}\), implying a pure diffusion whose generator is \(\tfrac{\sigma^2}{2}\,\Delta\). This diffusion is ergodic and converges to the uniform distribution over \(\mathcal{M}\) as \(t\to\infty\), reflecting the finite volume and absence of drift, so that the random walk explores all regions equally over long times.

For the simpler, yet illustrative, case \(\mathbf{b}(\mathbf{x},t)=f(t)\,\mathbf{x}\) and \(\sigma(\mathbf{x},t)=g(t)\) in \(\mathbb{R}^d\), the closed-form solution
\begin{equation}\label{eq:FP_ExactSol}
    p(\mathbf{x},t\mid \mathbf{x}_0)
    \;=\;
    \frac{1}{\bigl[2\pi\,\alpha^2(t)\,\beta^2(t)\bigr]^{\tfrac{d}{2}}}
    \exp\left[
       -\,\frac{\|\mathbf{x}-\alpha(t)\,\mathbf{x}_0\|^2}
              {\,2\,\alpha^2(t)\,\beta^2(t)}
    \right]
\end{equation}
arises, where
\begin{equation}\label{eq:scheduler}
   \alpha(t)
   \;=\; \exp\!\Bigl(\int_0^t f(t')\,dt'\Bigr),
   \quad
   \beta(t)
   \;=\; \sqrt{\int_0^t \frac{g^2(t')}{\alpha^2(t')}\,dt'}.
\end{equation}
If instead \(\mathbf{b}(\mathbf{x},t)\) is chosen as \(-\,\nabla U(\mathbf{x})\) or \(-\,\lambda\,\mathbf{x}\), then the late-time stationary distribution becomes a finite-variance Gaussian, akin to the Ornstein–Uhlenbeck (OU) process. In a more general manifold setting, local expansions or approximate solutions can still guide how \(\mathbf{x}_t\) evolves, and one obtains an analogous \(\nabla_{\mathbf{x}}\!\log p(\mathbf{x},t\mid\mathbf{x}_0)\) for the \emph{reverse} SDE.

\subsection{Linear Scheduler for the Diffusion Coefficient}
\label{sec:scheduler}
A common design choice in \emph{score-based} or \emph{diffusion-based} models is to let the diffusion amplitude \(\sigma(\mathbf{x},t)\) grow over time in a simple, \emph{linear} fashion.  Denoting the squared diffusion coefficient by \(g^2(t)\), we set
\begin{equation}
    \label{eq:linear_scheduler}
    \sigma^2(\mathbf{x},t) 
    \;\equiv\; 
    g^2(t)
    \;=\;
    h_0
    \;+\;
    \bigl(h_f - h_0\bigr)\,\frac{t}{T},
\end{equation}
where \(h_0,h_f>0\) are chosen so that the noise is mild at \(t=0\) (ensuring minimal corruption near the start) and large at \(t=T\) (ensuring a wide or uniform final distribution).  On a \emph{compact} manifold, large $g^2(T)$ promotes \emph{uniform} coverage, whereas on a \emph{non-compact} manifold (e.g.\ Euclidean-like domains), it broadens the distribution similarly to a Gaussian random walk.  

Table~\ref{tab:select_b} summarizes our choice of $(h_0,h_f)$ for manifolds such as a 3D torus ($\mathbb{T}^3$), a hypercube ($\mathbb{C}^d$), or Euclidean space ($\mathbb{R}^{3 \times 3}$). 

\begin{table}[h]
\centering
\caption{Our choice of $(h_0,h_f)$ for the linear scheduler Eq.~\eqref{eq:linear_scheduler} on different manifolds.}
\begin{ruledtabular}
\begin{tabular}{ccc}
    Manifold & $h_{0}$ & $h_{f}$ \\
    \hline
    $\mathbb{T}^{3}$ & $10^{-4}$ & 1 \\
    $\mathbb{C}^{d}$ & $10^{-6}$ & 5 \\
    $\mathbb{R}^{3 \times 3}$ & $10^{-3}$ & 20
\end{tabular}
\end{ruledtabular}
\label{tab:select_b}
\end{table}

\begin{figure*}[ht!]
    \centering
    \includegraphics[width=0.76\linewidth]{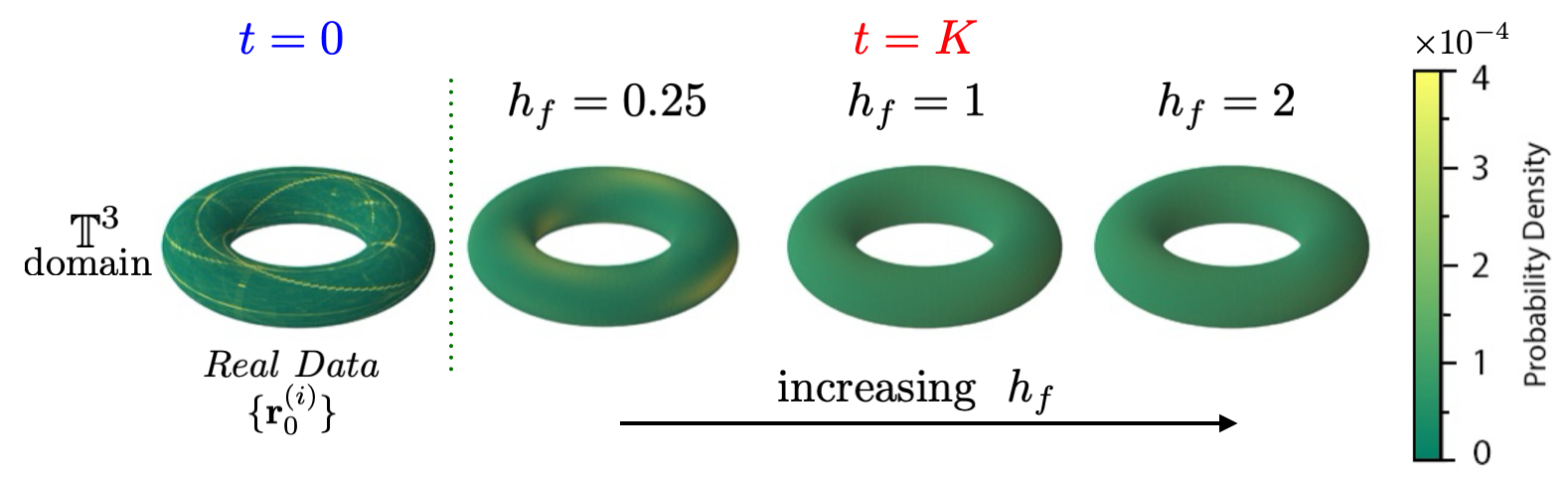}
    \caption{
    Final random-walk distributions on the 3D torus \(\mathbb{T}^3\) for various \(h_f\) values. The leftmost figure illustrates the initial distribution of real data, with each data point to evolve and be sampled 1000 times. As \(h_f\) increases from left to right (\(h_f = 0.25, 1, 2\)), the random walk becomes more uniform, though excessively large \(h_f\) can degrade training stability. The late-time distribution saturates the uniform density at $h_f = 2$.
    }
    \label{fig:torus_bf}
\end{figure*}

\begin{figure*}[ht!]
    \centering
    \includegraphics[width=0.75\linewidth]{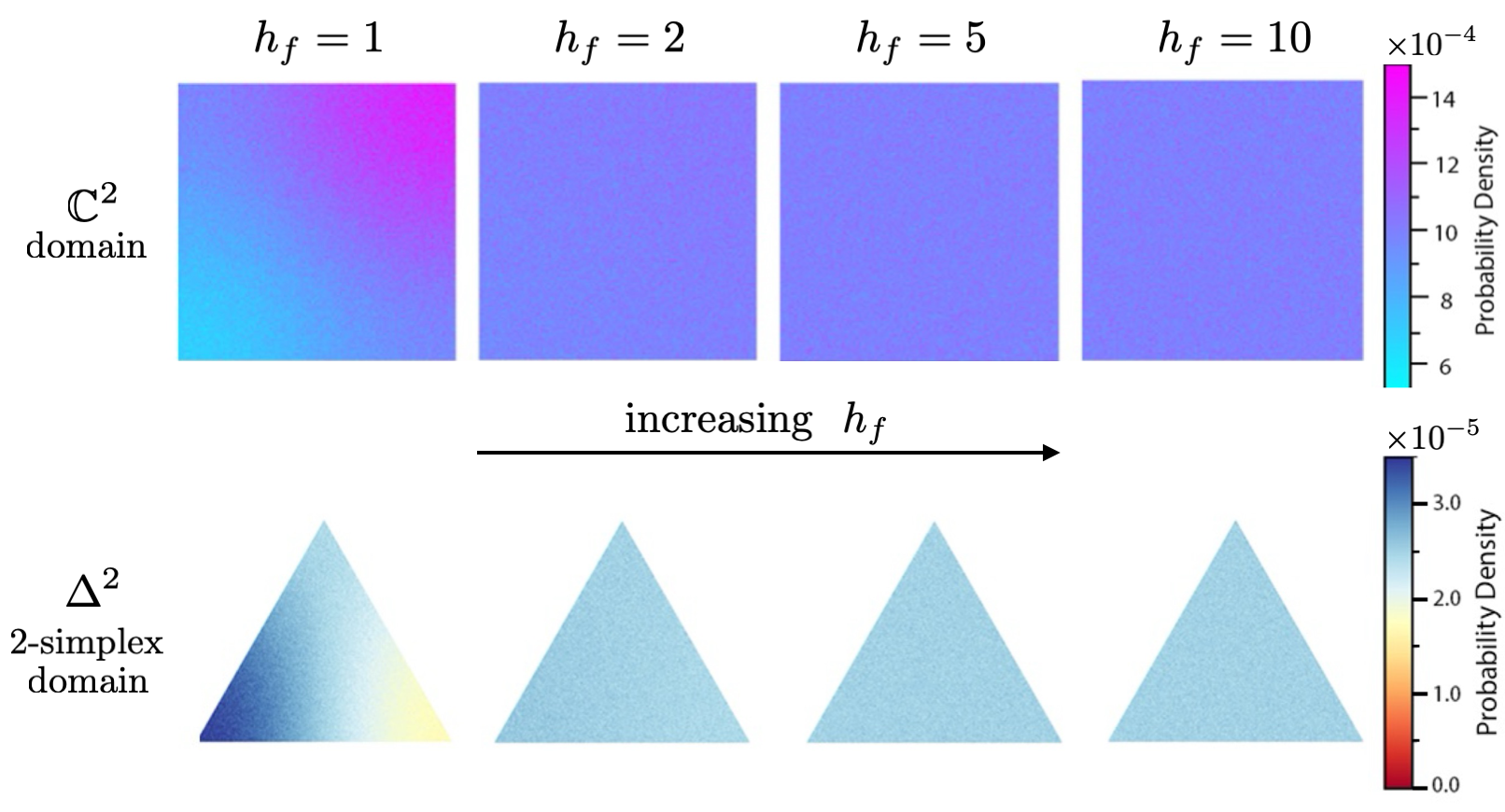}
   \caption{Top row: Final random-walk distributions on a 2D hypercube $\mathbb{C}^2 = [0,1]^2$ for various $h_f$ values, each starting from a single point and sampled $5\times 10^{7}$ times. 
Bottom row: The corresponding probability densities on the 2-simplex $\Delta^2$, obtained by applying the hypercube spacing transform to the top-row distributions. 
As $h_f$ increases from 1 to 10 (left to right), the random walk becomes increasingly uniform and saturates the uniform density at $h_f = 5$ in both $\mathbb{C}^2$ and its simplex representation.}
    \label{fig:hpc_simp_bf}
\end{figure*}

Fig.~\ref{fig:torus_bf} illustrates how varying $h_f$ affects the final random-walk distribution on a torus, with each run starting from points in the dataset. As $h_f$ increases, the distribution becomes more uniform; however, experiments show that both training and generation quality degrade significantly once $h_f$ is too large. In practice, $h_f=1$ gives a good balance that preserves the data distribution while still provides enough diffusion. For the hypercube manifold, setting $h_f=5$ similarly yields a sufficiently uniform distribution and facilitates the mapping to the uniform distribution on the simplex (see Fig.~\ref{fig:hpc_simp_bf}).

\section{Adaptive Timestep and No-Drift Approximation}
\label{sec:adaptive_t}

\paragraph{Adaptive step size.}
Numerically solving the reverse-time SDE (Section~\ref{sec:RSGM}) often involves discretizing time and applying a geodesic random walk step at each interval. A simple fixed-step scheme (e.g. Euler-Maruyama) can be suboptimal, so we use an \emph{adaptive} timestep, which has been described in earlier publications  \cite{DBLP:journals/corr/abs-2201-06503} and \cite{DBLP:journals/corr/abs-2106-03802}, by choosing
\[
   \gamma(t) \;=\; t^\xi\,\frac{T}{K},
\]
where $t$ is the current time in reverse diffusion, $T$ is the final time, $K$ is the total number of timesteps, and $\xi$ is a hyperparameter. Intuitively, $\gamma(t)$ shrinks at early times if $\xi>0$, allowing finer resolution when $\nabla_{\mathbf{x}}\log p_t(\mathbf{x})$ changes quickly, and grows at later times to speed up sampling. Empirically, we find that $\xi=1$ yields higher-quality samples compared to a constant $\Delta t$ approach, presumably because the network can better invert the forward noising at early timesteps.

\paragraph{No drift (variance-exploding) SDE in Euclidean space.}
For a Euclidean manifold $\mathbb{R}^d$, one may neglect the drift $\mathbf{b}(\mathbf{x},t)$, yielding a \emph{variance-exploding} process \cite{Song2021}. Namely, if $f(t)=0$ in Eq.~\eqref{eq:FP_ExactSol}, the diffusion amplitude $\beta^2(t)$ must be large enough at $t=T$ to overwhelm the original data distribution such that
\[
   p(\mathbf{x}_T\mid \mathbf{x}_0)
   \;=\;
   \mathcal{N}\!\bigl(\mathbf{x}_0,\,\beta^2(T)\bigr)
   \;\approx\;
   \mathcal{N}\!\bigl(\mathbf{0},\,\beta^2(T)\bigr),
\]
where we assume the effect of the original mean of the data is negligible once $\beta^2(T)$ becomes sufficiently large. In practice, ignoring the drift in $\mathbb{R}^d$ simplifies the forward noising yet still ensures that $p(\mathbf{x}_T)$ is effectively wide by time $T$.

\paragraph{One-step random walk vs.\ multiple small steps.}
A previous study \cite{Karras2022elucidating} shows that if the SDE drift is $f(t)\,\mathbf{x}$ and the diffusion $g(t)$ depends only on $t$, then the transition kernel simplifies to Eq.~\eqref{eq:FP_ExactSol}
\[
   p(\mathbf{x}_t\mid \mathbf{x}_0)
   \;=\;
   \mathcal{N}\Bigl(\alpha(t)\,\mathbf{x}_0,\;\alpha^2(t)\,\beta^2(t)\Bigr),
\]
where $\alpha(t)$ and $\beta(t)$ follow Eq.~\eqref{eq:scheduler}. This kernel can be integrated in \emph{one} geodesic random walk step per diffusion step, rather than many small increments, giving an equivalent result at a smaller computational cost. For instance, using the closed-form solution in Eq.~\eqref{eq:FP_ExactSol} directly replaces multiple fine steps by a single jump, and speed up training while preserving exactness in the forward or backward transitions.

In summary, for the implementation of geometric random walks, we utilize (1) the no-drift variance-exploding approach \cite{Song2021} for the random walks in $\mathbb{R}^d$, (2) a single random-walk increment ($K=1$) instead of many small increments during the training (Algorithm~\ref{algo:train_real}), and (3) \emph{adaptive} timesteps for discretizing the reverse SDE. These strategies yield higher-quality samples and improve computational efficiency.

\section{Stable, unique, and novel structures}
\label{sec:unique_novel_appendix}

\subsection{ALEX-MP dataset}
The dataset contains 845,995 materials, merged from Alexandria and Material Projects databases. It is bigger than ALEX-MP-20 dataset which is screened for materials with $E_{\textrm{hull}} < 100$ meV and the number of atoms $\leq$ 20 atoms per unit cell. However, MatterGen evaluates novelty using ALEX-MP-ICSD dataset, which combines ALEX-MP with 117.7K disordered structures from the ICSD database--access to which requires a license. To generate crystal structures, we first sample the number of atoms per unit cell ($N$) from its empirical distribution in the ALEX-MP-20 dataset. Consequently, for the 10$^4$ generated structures, the distribution of $N$ closely matches that of the training set (Fig.~\ref{fig:alexmp20_stats}(a)).

\subsection{Uniqueness and novelty}
The similarity between two structures can be evaluated using \texttt{StructureMatcher} from the \texttt{pymatgen}, with tolerance parameters set to \texttt{ltol=0.2, stol=0.3, angle\_tol=5.0}, following MatterGen's setting \cite{Zeni2025mattergen}. \texttt{StructureMatcher} receives a pair of structures with the same compositions and determines the similarity of their lattice vectors and atomic positions within the Niggli cell. It also considers permutations of atoms within each of the two cells. 

To compute uniqueness, the similarities between every pair of structures in the generated set are evaluated. The number of similar pairs is then divided by the total number of structures in the generation set. For novelty, each structure in the generation set is compared with every structure in the ALEX-MP dataset. The number of generated structures that are not similar to any structures in the ALEX-MP dataset is then divided by the total number of generated structures.

\subsection{Examples of generated structures}
\begin{figure*}[ht!]
    \centering
    \includegraphics[width=0.75\linewidth]{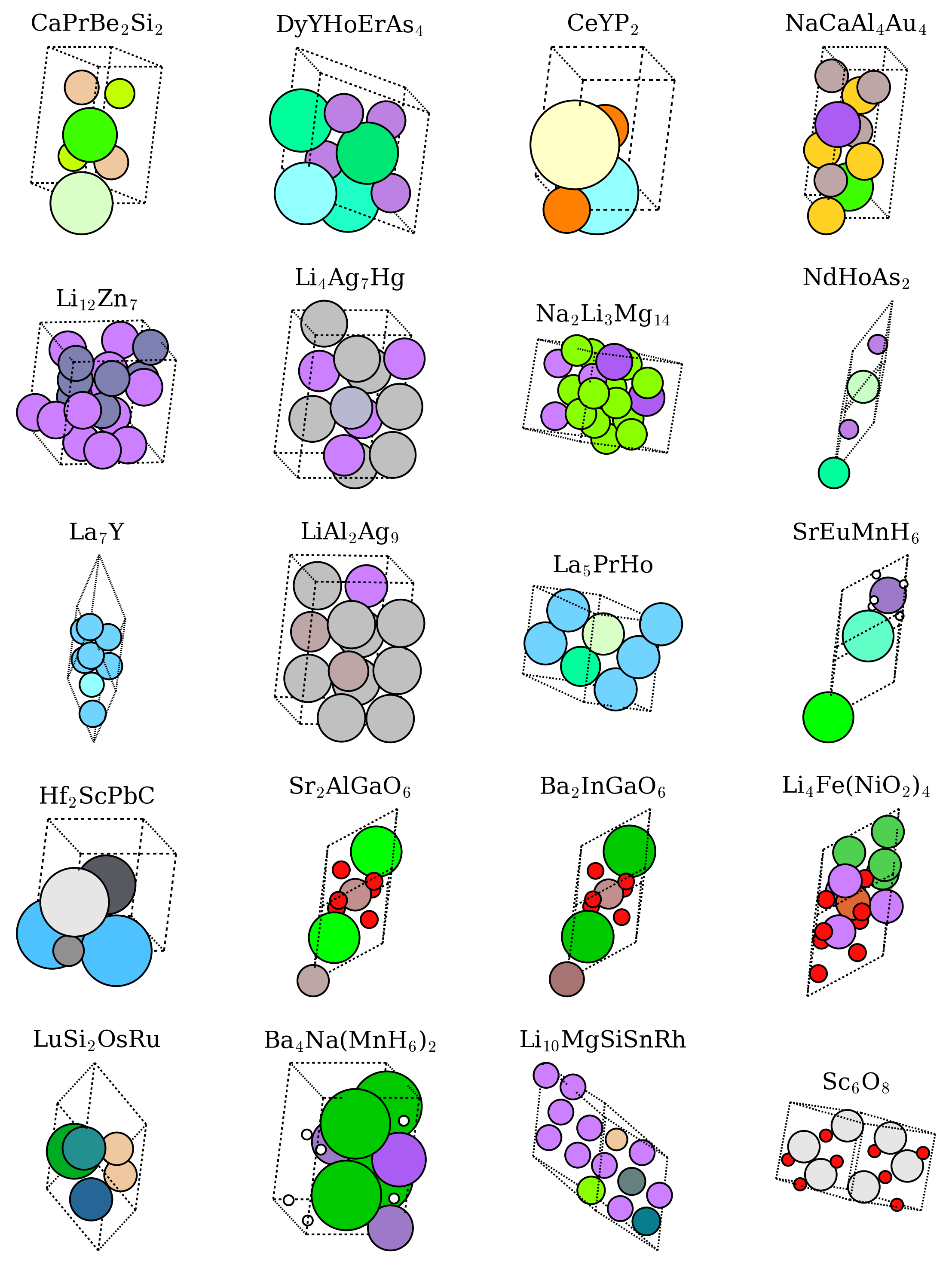}
   \caption{Novel structures presented in Table~\ref{tab:novel_structures}.}
    \label{fig:novel_structures}
\end{figure*}

Table~\ref{tab:novel_structures} presents example of S.U.N. structures, and Fig.~\ref{fig:novel_structures} illustrates S.U.N. structures reported in Table~\ref{tab:novel_structures}. Table~\ref{tab:high_rmsd_structures} shows generated structures that have RMSD $> 0.1$ {\AA}.

Table~\ref{tab:pg_percent_match} reports F1 scores, success rates, uniquenesses, and percentages occurrence in the training set for 32 point groups. We find that all generated structures are structurally valid, with 99.12\% and 96.78\% compositionally valid for 2 and 4 atoms per unit cell, respectively. Tables~\ref{tab:pg_2atoms} and \ref{tab:pg_4atoms} list point-group-controlled structures with 2 and 4 atoms per unit cell, respectively. 

\begin{table*}
\centering
\caption{Examples of novel materials that have $E_{\textrm{hull}} < 100$ meV.}
\begin{ruledtabular}
\begin{tabular}{cccccc}
    Formula & Space group & \# Atoms per unit cell & Charge balance & $E_{\textrm{hull}}$ (meV) & RMSD ({\AA}) \\
    \hline
    CaPrBe$_2$Si$_2$ & $P\bar{3}m1$ & 6 & Yes & 2.25 & 0.0088 \\
    DyYHoErAs$_4$ & $P2/m$ & 8 & Yes & 2.41 & 0.0010 \\
    CeYP$_2$ & $P4/mmm$ & 4 & Yes & 3.33 & 0.0  \\
    NaCaAl$_4$Au$_4$ & $P4mm$ & 10 & Yes & 4.35 & 0.0022 \\
    Li$_{12}$Zn$_7$ & $R\bar{3}m$ & 19 & Yes & 4.65 & 0.0057 \\
    Li$_4$Ag$_7$Hg & $Pm$ & 12 & Yes & 7.57 & 0.0184 \\
    Na$_2$Li$_3$Mn$_{14}$ & $P1$ & 19 & Yes & 7.70 & 0.0121 \\
    NdHoAs$_2$ & $R\bar{3}m$ & 4 & Yes & 8.43 &  3.9$\times 10^{-5}$ \\
    La$_7$Y & $I4/mmm$ & 8 & Yes & 8.67 & 0.0074  \\
    LiAl$_2$Ag$_9$ & $Pm$ & 12 & Yes & 9.41 & 0.0260 \\
    La$_5$PrHo & $Cm$ & 7 & Yes & 9.75 & 0.0017 \\
    SrEuMnH$_6$ & $F\bar{4}3m$ & 9 & Yes & 16.75 & 0.0015 \\
    Hf$_2$ScPbC & $P4/mmm$ & 5 & Yes & 49.43 & 0.0 \\
    Sr$_2$AlGaO$_6$ & $Fm\bar{3}m$ & 10 & Yes & 54.05 & 0.0005 \\
    Ba$_2$InGaO$_6$ & $Fm\bar{3}m$ & 10 & Yes & 57.46 & 0.0180 \\
    Li$_4$Fe(NiO$_2$)$_4$ & $F\bar{4}3m$ & 17 & Yes & 74.17 & 0.0703 \\

    LuSi$_2$OsRu & $I\bar{4}m2$ & 5 & No & 5.26 & 0.0032 \\
    Ba$_4$Na(MnH$_6$)$_2$ & $P4/mmm$ & 19 & No & 8.35 & 0.0341 \\
    Li$_{10}$MgSiSnRh & $Cm$ & 14 & No & 9.55 & 0.0356 \\

\end{tabular}
\end{ruledtabular}
\label{tab:novel_structures}
\end{table*}

\begin{table*}
\centering
\caption{Generated materials with RMSD $> 0.1$ {\AA}.}
\begin{ruledtabular}
\begin{tabular}{cccccc}
    Formula & Space group & \# Atoms per unit cell & Charge balance & $E_{\textrm{hull}}$ (meV) & RMSD ({\AA}) \\
    \hline
    Li$_5$Fe$_3$NiBF$_7$ & $Cm$ & 17 & Yes & 92.13 & 0.268 \\
    Li$_2$CuNi$_9$BRhO$_5$ & $P1$ & 19 & Yes & 137.93 & 0.233 \\
    Ni$_13$O$_6$F & $Cm$ & 20 & Yes & 144.94 & 0.242 \\
    Li$_3$Mn$_2$FeNi$_5$O$_8$ & $Cm$ & 19 & Yes & 93.25 & 0.156 \\
    VZnCu$_9$NiGeH & $P1$ & 14 & No & 36.48 & 0.111 \\
    Ti$_8$VO$_9$ & $C2/m$ & 18 & Yes & 41.69 & 0.114 \\
    LiMn$_2$Ni$_7$O$_9$ & $Cm$ & 19 & Yes & 106.90 & 0.213 \\
    Sc$_3$O$_4$ & $C2/m$ & 14 & No & 48.86 & 0.1890 \\

\end{tabular}
\end{ruledtabular}
\label{tab:high_rmsd_structures}
\end{table*}

\begin{table*}
\centering
\caption{Macro F1 scores and success rates are computed for 32 point groups and compared to each point group's presence in the training set for 2 and 4 atoms per unit cell cases. The final column reports their overall existence in the full training set. Uniquenesses are computed over the structures whose point group symmetry matches the input point group.}
\begin{ruledtabular}
\begin{tabular}{cccccccccc}
    Point group & \multicolumn{4}{c}{2 atoms per unit cell} & \multicolumn{4}{c}{4 atoms per unit cell} & Full training set (\%) \\
    & F1 & Success rate & Training set & Unique & F1 & Success rate & Training set & Unique \\
    \hline
    $1$ & 0.015 & 0.78 (1/128) & 0.54 & 100 & 0.043 & 2.34 & 0.27 & 100 & 1.26 \\
    $\bar{1}$ & 0.304 & 26.56 & 0.95 & 58.82 & 0.199 & 16.4 & 0.18 & 66.67 & 2.20 \\
    $2$ & 0.0 & 0.0 & 0.0 & -- & 0.059 & 3.13 & 0.11 & 100 & 0.51 \\
    $m$ & 0.0 & 0.0 & 0.77 & -- & 0.505 & 37.50 & 2.85 & 85.42 & 1.31 \\
    $3$ & 0.0 & 0.0 & 0.001 & -- & 0.0 & 0.0 & 0.0 & -- & 0.94 \\
    $2/m$ & 0.349 & 82.03 & 2.68 & 40.95 & 0.411 & 57.03 & 2.7 & 68.49 & 23.85 \\
    $222$ & 0.031 & 1.56 & 0.06 & 100 & 0.0 & 0.0 & 0.02 & -- & 0.16 \\
    $mm2$ & 0.133 & 9.38 & 1.31 & 100 & 0.832 & 89.06 & 6.91 & 95.61 & 4.57 \\
    $4$ & 0.0 & 0.0 & 0.0 & -- & 0.0 & 0.0 & 0.0 & -- & 0.06 \\
    $\bar{4}$ & 0.0 & 0.0 & 0.0 & -- & 0.0 & 0.0 & 0.0 & -- & 0.40 \\
    $\bar{3}$ & 0.0 & 0.0 & 0.0 & -- & 0.0 & 0.0 & 0.0 & -- & 1.47 \\
    $32$ & 0.0 & 0.0 & 0.0 & -- & 0.0 & 0.0 & 0.002 & -- & 0.78 \\
    $3m$ & 0.045 & 2.34 & 1.01 & 100 & 0.590 & 100 & 7.78 & 87.50 & 2.34 \\
    $6$ & 0.0 & 0.0 & 0.0 & -- & 0.0 & 0.0 & 0.0 & -- & 0.003 \\
    $\bar{6}$ & 0.0 & 0.0 & 0.0 & -- & 0.0 & 0.0 & 0.0 & -- & 0.005 \\
    $mmm$ & 0.431 & 63.28 & 7.08 & 54.32 & 0.306 & 51.56 & 6.44 & 98.48 & 12.01 \\
    $4/m$ & 0.0 & 0.0 & 0.0 & -- & 0.0 & 0.0 & 0.0 & -- & 0.30 \\
    $422$ & 0.0 & 0.0 & 0.0 & -- & 0.0 & 0.0 & 0.0 & -- & 0.03 \\
    $4mm$ & 0.293 & 18.75 & 0.54 & 66.67 & 0.928 & 100 & 6.02 & 90.63 & 2.06 \\
    $\bar{4}2m$ & 0.258 & 32.03 & 0.48 & 73.17 & 0.391 & 61.72 & 1.76 & 96.20 & 4.69 \\
    $\bar{3}m$ & 0.419 & 80.47 & 5.77 & 59.22 & 0.366 & 79.69 & 8.58 & 86.27 & 11.43 \\
    $6/m$ & 0.0 & 0.0 & 0.0 & -- & 0.0 & 0.0 & 0.0 & -- & 0.15 \\
    $622$ & 0.0 & 0.0 & 0.0 & -- & 0.0 & 0.0 & 0.0 & -- & 0.025 \\
    $6mm$ & 0.0 & 0.0 & 0.0 & -- & 0.429 & 29.69 & 0.085 & 26.32 & 0.15  \\
    $\bar{6}m2$ & 0.409 & 75.78 & 10.95 & 70.10 & 0.251 & 25.78 & 0.36 & 78.79 & 2.32 \\
    $23$ & 0.0 & 0.0 & 0.0 & -- & 0.0 & 0.0 & 0.0 & -- & 0.08 \\
    $4/mmm$ & 0.515 & 94.53 & 16.60 & 78.51 & 0.272 & 71.09 & 28.28 & 100 & 16.32 \\
    $6/mmm$ & 0.251 & 82.81 & 2.86 & 13.21 & 0.356 & 93.75 & 0.38 & 51.67 & 2.31 \\
    $m\bar{3}$ & 0.0 & 0.0 & 0.0 & -- & 0.0 & 0.0 & 0.0 & -- & 0.23 \\
    $432$ & 0.0 & 0.0 & 0.0 & -- & 0.0 & 0.0 & 0.0 & -- & 0.002 \\
    $\bar{4}3m$ & 0.450 & 77.34 & 5.41 & 54.55 & 0.625 & 86.72 & 17.58 & 98.20 & 4.13 \\
    $m\bar{3}m$ & 0.249 & 100 & 43.01 & 89.84 & 0.339 & 95.31 & 9.70 & 99.18 & 3.90 \\
\end{tabular}

\end{ruledtabular}
\label{tab:pg_percent_match}
\end{table*}

\begin{figure*}[ht!]
    \centering
    \includegraphics[width=0.75\linewidth]{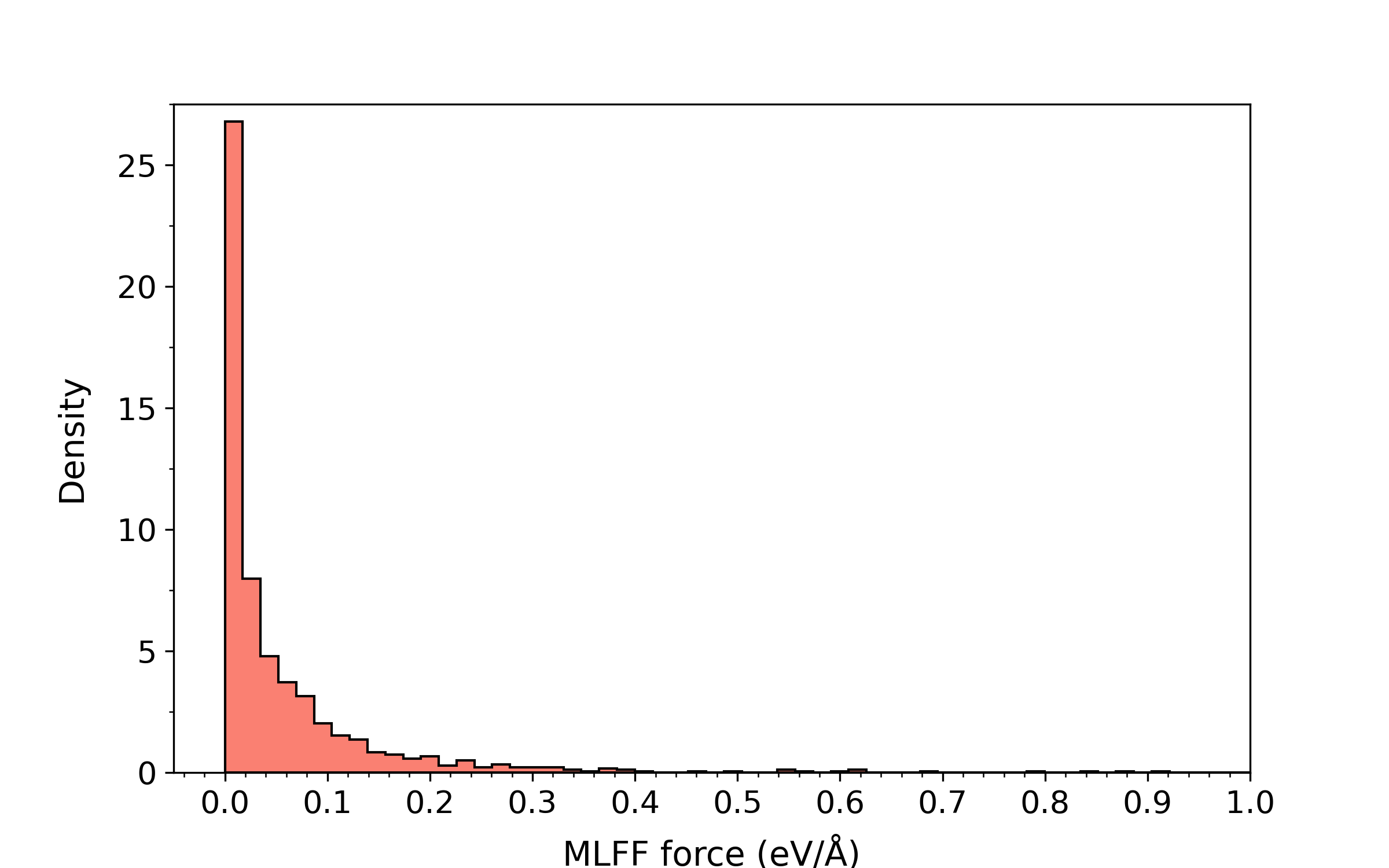}
   \caption{Distribution of average force predictions using SevenNet \cite{Park2024sevennet} in the unit of eV/{\AA}.}
    \label{fig:mlff_force}
\end{figure*}

\begin{table}
\centering
\caption{Examples of generated structures with space groups correctly controlled by the input point group for cases with 2 atoms per unit cell.}
\begin{ruledtabular}
\begin{tabular}{ccc}
    Input point group & Output space group & Formula \\
    \hline \\[0.01em]
    $\bar{1}$ & $P\bar{1}$ & Ga \\
    $2/m$ & $C2/m$ & MgCd \\
    $mm2$ & $Imm2$ & NiAs \\
    $222$ & $F222$ & VH \\
    $mm2$ & $Imm2$ & GaPb \\
     & $Amm2$ & CdAg \\
    $3m$ & $R3m$ & BiS \\
    $mmm$ & $Cmmm$ & ZnCu \\
    $4mm$ & $I4mm$ & SnSb \\
    $\bar{4}2m$ & $I\bar{4}m2$ & CdO \\
    $\bar{3}m$ & $R\bar{3}m$ & CoIr \\
    $\bar{6}m2$ & $P\bar{6}m2$ & CoNi \\
    $4/mmm$ & $P4/mmm$ & AgAu \\
     & $I4/mmm$ & Zn \\
    $6/mmm$ & $P6_3/mmc$ & Mn \\
    $\bar{4}3m$ & $F\bar{4}3m$ & CsF \\
    $m\bar{3}m$ & $Pm\bar{3}m$ & MgZn \\
     & $Fm\bar{3}m$ & SbPb \\
\end{tabular}

\end{ruledtabular}
\label{tab:pg_2atoms}
\end{table}

\begin{table}
\centering
\caption{Examples of generated structures with space groups correctly controlled by the input point group for cases with 4 atoms per unit cell.}
\begin{ruledtabular}
\begin{tabular}{ccc}
    Input point group & Output space group & Formula \\
    \hline
    $1$ & $P1$ & InSnHg$_2$ \\
    $\bar{1}$ & $P\bar{1}$ & Al$_3$Ga \\
    $2$ & $C2$ & Tl$_2$InPb \\
    $m$ & $Cm$ & FeNiRh$_2$ \\
    $2/m$ & $C2/m$ & MnNi$_3$ \\
    $mm2$ & $Amm2$ & TbCeGaGe \\
     & $Pmm2$ & Cu$_3$Ni \\
    $3m$ & $R3m$ & KZrS$_2$ \\
    $mmm$ & $Cmmm$ & FeNi$_3$ \\
     & $Pmmm$ & SrMgInPb \\
     & $Immm$ & GaFe$_2$Co \\
    $4mm$ & $P4mm$ & LaHoTlAu \\
    $\bar{4}2m$ & $I\bar{4}m2$ & CuPt$_2$Rh \\
    $\bar{3}m$ & $R\bar{3}m$ & PrYP$_2$ \\
    $6mm$ & $P6_3mc$ & ZnSe \\
    $\bar{6}m2$ & $P\bar{6}m2$ & CoNi$_3$ \\
    $4/mmm$ & $P4/mmm$ & SmDy \\
    $6/mmm$ & $P6/mmm$ & Sn \\
     & $P6_3/mmc$ & YTe \\
    $\bar{4}3m$ & $F\bar{4}3m$ & CaYZnIn \\
    $m\bar{3}m$ & $Pm\bar{3}m$ & CeHg \\
     & $Fm\bar{3}m$ & TaMo$_3$ \\
\end{tabular}

\end{ruledtabular}
\label{tab:pg_4atoms}
\end{table}

\section{Hyperparameters}
\label{sec:hyperparameters_appendix}

We train our models using a learning-rate scheduler that reduces the learning rate upon detecting plateaus in training loss. Specifically, we begin with a maximum learning rate of $4\times 10^{-4}$, apply a patience of 30 epochs (with a 0.6 reduction factor), and allow up to 300 epochs of patience before stopping. Unless otherwise noted, the batch size is 32 for experiments on the MP-20 and ALEX-MP-20 datasets. Our equivariant GNN follows the 31M EquiformerV2 model described in \cite{Liao2024equiformerv2}, with the key hyperparameters summarized in Table~\ref{tab:hyperparameters}.

\begin{table*}[ht!]
\centering
\caption{Key hyperparameters for CrystalGRW.}
\begin{ruledtabular}
\begin{tabular}{lc}
\textbf{Hyperparameters} & \\
\hline
\\
\underline{Optimizer \& Scheduler} \\
Optimizer & AdamW \\
Learning-rate scheduling & Reduce on plateau \\
Max learning rate & $4 \times 10^{-4}$ \\
Weight decay & $1 \times 10^{-3}$ \\
Model EMA decay & 0.999 \\
Gradient clipping norm threshold & 100 \\
Batch size & 32 \\
\\
\underline{Loss Weights}\\
Coordinate weight $\lambda_{(\mathbb{T}^3)^N}$ & 1 \\
Atom-type weight $\lambda_{(\mathbb{C}^d)^N}$ & 1 \\
Lattice weight $\lambda_{\mathbb{R}^{3 \times 3}}$ & 1 \\
\\
\underline{EquiformerV2 Setup}\\
Cutoff radius (\AA) & 12 \\
Maximum number of neighbors & 12 \\
Number of radial bases & 512 \\
$d_{\text{edge}}$ (hidden scalar dimension for radial functions) & (0,\,128) \\
Maximum degree $l_{\max}$ & 4 \\
Maximum order $m_{\max}$ & 2 \\
Number of Transformer blocks & 8 \\
Embedding dimension $d_{\text{embed}}$ & (4,\,128) \\
Attention hidden dimension $d_{\text{attn\_hidden}}$ & (4,\,64) \\
Number of attention heads $h$ & 8 \\
Attention alpha dimension $d_{\text{attn\_alpha}}$ & (0,\,64) \\
Attention value dimension $d_{\text{attn\_value}}$ & (4,\,16) \\
Feedforward network hidden dimension $d_{\text{ffn}}$ & (4,\,128) \\
Resolution of point samples $R$ & 18 \\
Dropout rate & 0.1 \\
\\
\underline{Output Heads} \\
Coordinate hidden dimension & 128 \\
Atom-type hidden dimension & 128 \\
Lattice hidden dimension & 256 \\
\\
\underline{Additional Features} \\
Time embedding dimension & 128 \\
Condition embedding dimension & 128 \\
Probability of training under the null condition (for training with guided conditions) & 0.1 \\
\\
\underline{Sampling parameters} \\
Total timesteps $K$ & 1000 \\
Final time $T$ & 1 \\
Adaptive timestep parameter $\xi$ & 1 \\
Guidance strength $w$ (for condition-guided generations) & 0.5 \\
\end{tabular}
\end{ruledtabular}
\label{tab:hyperparameters}
\end{table*}

\end{document}